\documentclass[12pt]{article}
\usepackage{amsmath,amssymb,amsthm,amsxtra,bbm,overpic,bm,epsfig,subfigure,tikz,tikz-feynman}
\usepackage{hyperref}
\usepackage{mathrsfs}
\usepackage{enumitem}
\usepackage{graphicx}
\usepackage{color}
\usepackage{comment}
\usepackage{epstopdf}
\usepackage{float}
\usepackage{cite}
\allowdisplaybreaks[2]
\usetikzlibrary{arrows.meta}
\makeatletter
\tikzfeynmanset{compat=1.1.0} 
\makeatother
\usepackage[all]{xy}

\usepackage{slashed,stmaryrd,multirow}
\numberwithin{equation}{section}

\def\thefootnote{\fnsymbol{footnote}}

\usepackage{hyperref}
\usepackage{slashed,stmaryrd,orcidlink}

\usepackage{lscape}%
\usepackage{array}
\usepackage{booktabs}%

\begin{document}
	
	\vspace{0.2cm}
	
	\begin{center}
		{\Large\bf Revealing precision bounds on neutrino oscillation parameters with quantum estimation theory}
	\end{center}
	
	\vspace{0.2cm}
	
	\begin{center}
		{\bf Jihong Huang}{\orcidlink{0000-0002-5092-7002}},$^{1,2}$\footnote{E-mail: huangjh@ihep.ac.cn}
		\ 
		{\bf Tommy Ohlsson}{\orcidlink{0000-0002-3525-8349}},$^{3,4}$\footnote{E-mail: tohlsson@kth.se}
		\ 
		{\bf Sampsa Vihonen}{\orcidlink{0000-0001-7761-2847}},$^{3,4}$\footnote{E-mail: vihonen@kth.se}
		\ 
		{\bf Shun Zhou}{\orcidlink{0000-0003-4572-9666}},$^{1,2}$\footnote{E-mail: zhoush@ihep.ac.cn}
		\\
		\vspace{0.2cm}
		{\small
			$^{1}$Institute of High Energy Physics, Chinese Academy of Sciences, Beijing 100049, China\\
			$^{2}$School of Physical Sciences, University of Chinese Academy of Sciences, Beijing 100049, China\\
			$^{3}$Department of Physics, School of Engineering Sciences, KTH Royal Institute of Technology,\\
			AlbaNova University Center, Roslagstullsbacken 21, SE–106 91 Stockholm, Sweden\\
			$^{4}$The Oskar Klein Centre, AlbaNova University Center, Roslagstullsbacken 21,\\
			SE–106 91 Stockholm, Sweden}
	\end{center}

	\vspace{0.5cm}
	
	\begin{abstract}
		Quantum estimation theory provides ultimate precision bounds on parameter estimation, independent of experimental setups. 
		In this article, we apply this theoretical framework to neutrino oscillations, aiming to clarify some subtle issues and reveal the maximum achievable precision of oscillation parameters. First, taking the example of two-flavor oscillations, we clarify how the quantum Fisher information (QFI) depends on the choice of bases when the basis transformation itself involves the parameters in question. Then, for three-flavor oscillations, we compute the QFI matrix for electron and muon neutrino states in the flavor basis and derive analytical expressions and numerical results for both diagonal and off-diagonal elements. The implications of off-diagonal correlations for multiparameter estimation are discussed, including the compatibility of optimal measurements, the Holevo bound, and the spectral analysis of the QFI matrix, from which we show that a single neutrino channel leaves exactly two parameter combinations unconstrained. The quantum Cram\'{e}r-Rao bounds on the precision of oscillation parameters for typical reactor and long-baseline accelerator neutrino experiments are obtained as vacuum, state-level benchmarks, which establish a theoretical reference for the ultimate precision achievable in future neutrino oscillation experiments.
	\end{abstract}

	\def\thefootnote{\arabic{footnote}}
	\setcounter{footnote}{0}

	\newpage
	
	\section{Introduction}
	
	The experimental discovery of neutrino oscillations indicates that neutrinos are massive and the flavor mixing in the lepton sector is significant. 
	In the standard parametrization advocated by the Particle Data Group~\cite{ParticleDataGroup:2026aaa}, the Pontecorvo-Maki-Nakagawa-Sakata (PMNS) matrix~\cite{Pontecorvo:1957cp,Maki:1962mu} describing the leptonic flavor mixing is expressed as
	\begin{eqnarray}
		\label{eq:PMNS}
		U = \begin{pmatrix}
			c_{12}^{} c_{13}^{} & s_{12}^{} c_{13}^{} & s_{13}^{} {\rm e}^{-{\rm i} \delta_{\rm CP}^{}}  \\
			-s_{12}^{} c_{23}^{} - c_{12}^{} s_{13}^{} s_{23}^{} {\rm e}^{{\rm i}\delta_{\rm CP}^{}} & c_{12}^{} c_{23}^{} - s_{12}^{} s_{13}^{} s_{23}^{} {\rm e}^{{\rm i} \delta_{\rm CP}^{}} & c_{13}^{} s_{23}^{} \\
			s_{12}^{} s_{23}^{} - c_{12}^{} s_{13}^{} c_{23}^{} {\rm e}^{{\rm i}\delta_{\rm CP}^{}} & -c_{12}^{} s_{23}^{} - s_{12}^{} s_{13}^{} c_{23}^{} {\rm e}^{{\rm i} \delta_{\rm CP}^{}} & c_{13}^{} c_{23}^{} 
		\end{pmatrix} \cdot P_{\nu}^{} \;,
	\end{eqnarray}
	with $c_{ij}^{} \equiv \cos\theta_{ij}$ and $s_{ij}^{} \equiv \sin\theta_{ij}^{}$ (for $i,j=1,2;1,3;2,3$). 
	The three mixing angles $\{\theta_{12}^{}, \theta_{13}^{}, \theta_{23}^{}\}$ and the Dirac {\it CP}-violating phase $\delta_{\rm CP}^{}$ are directly relevant to neutrino oscillation probabilities, while the two Majorana phases contained in the diagonal matrix $P_\nu^{}$ are irrelevant. In addition, neutrino oscillation probabilities depend on two mass-squared differences $\Delta m_{21}^2 \equiv m_2^2 - m_1^2$ and $ \Delta m_{31}^2 \equiv m_3^2 - m_1^2$.
 The current and future neutrino oscillation experiments, including the reactor neutrino experiment JUNO~\cite{JUNO:2015zny} and long-baseline accelerator neutrino experiments (such as T2HK~\cite{Hyper-Kamiokande:2018ofw}, DUNE~\cite{DUNE:2020ypp} and ESSnuSB~\cite{Alekou:2022emd}), are dedicated to determining whether the neutrino masses follow the normal ordering (NO, $m_1^{} < m_2^{} < m_3^{}$) or the inverted ordering (IO, $m_3^{} < m_1^{} < m_2^{}$), and to searching for leptonic {\it CP} violation. 
In addition, these experiments will perform precise measurements of oscillation parameters with subpercent accuracy~\cite{JUNO:2022mxj,Capozzi:2025wyn,JUNO:2025gmd}. 
	
In the era of precision measurements, there arises a natural but important question: What precision can experiments achieve in measuring the neutrino oscillation parameters? Quantum estimation theory (QET) provides a rigorous theoretical framework for answering this question~\cite{Helstrom:1969fri}, by introducing the quantum Fisher information (QFI) that quantifies how sensitively a quantum state depends on an unknown physical parameter~\cite{Helstrom:1967ldp}.
	It generalizes the concept of the classical Fisher information, and sets the fundamental bound on the variance of any unbiased estimator through the quantum Cram\'{e}r-Rao (CR) inequality~\cite{Cramer1946,Rao1945}
	\begin{eqnarray}
		\mathrm{Var}[\lambda] \geqslant \frac{1}{N Q_\lambda^{}} \;,
	\end{eqnarray}
	which states that the estimation uncertainty of the parameter $\lambda$ is inversely proportional to the corresponding QFI, denoted as $Q_\lambda^{}$, and the number of experimental repetitions $N$.
	This serves as an important reference for neutrino oscillation experiments and can be used to maximize the accuracy of measurements.
	
	As central concepts in QET and key factors in experimental precision, the QFI and the CR bound have recently drawn significant attention. In neutrino physics, the QFI has been utilized to discuss the precision of oscillation parameter measurements in future experiments~\cite{Nogueira:2016qsk,Ignoti:2025rxr,Yadav:2026lsx,Chundawat:2026jjd,Frugiuele:2026yeq,Farooq:2026eap,Chundawat:2026lcm}. For example, Refs.~\cite{Yadav:2026lsx,Chundawat:2026lcm} show that the QFI can quantify the sensitivity of long-baseline neutrino experiments to oscillation parameters, providing a systematic way to identify which parameters are most accessible to measurement. Similarly, Ref.~\cite{Chundawat:2026jjd} investigates information-theoretic limitations in solar and reactor neutrino experiments, revealing that current experimental setups may not extract all available information about oscillation parameters. References~\cite{Ignoti:2025rxr,Frugiuele:2026yeq} employ the QFI to derive fundamental precision bounds for neutrino oscillation measurements, clarifying how quantum statistical limits constrain achievable sensitivities in realistic experiments in a complementary way. With a more experimental emphasis, quantum-information measures have also been applied to the sensitivity of concrete long-baseline experiments~\cite{Banerjee:2024lih,Dixit:2026qin}. These developments highlight how the QFI provides a powerful bridge between QET and neutrino physics, offering a rigorous framework for evaluating and optimizing the precision of future neutrino oscillation experiments.
    
    Nevertheless, the existing work on the QFI and the CR bound in neutrino physics remains rather limited. 
    Unlike other quantum systems, neutrino oscillations involve the transformation between two different types of eigenstates, namely, neutrino mass eigenstates and flavor eigenstates.
    The neutrino oscillation parameters of interest, such as the mixing angles $\theta_{ij}^{}$ and the {\it CP}-violating phase $\delta_{\rm CP}^{}$, are exactly those parameters embedded in the corresponding unitary transformation matrix, i.e., the PMNS matrix $U$. 
    For this reason, we first discuss how the QFI changes when quantum states are represented in different physical bases. 
    We consider two distinct cases, in which the basis transformation does or does not depend on the parameters to be estimated.
    Taking two-flavor neutrino oscillations as an example, we compute the QFI on the mixing angle and the mass-squared difference in both the mass basis and the flavor basis. The difference in the QFI is explicitly demonstrated. Then, given that neutrinos are actually produced and detected in the flavor basis, we calculate the QFI for three-flavor neutrino oscillations in the flavor basis. For next-generation neutrino oscillation experiments, we investigate the QFI on various parameters extracted from $\nu_e^{}$ and $\nu_\mu^{}$ beams, providing both analytical expressions with proper approximations and the full numerical results. Notably, we discuss the QFI with multiparameter estimation for the first time in neutrino physics, i.e., the off-diagonal elements of the QFI matrix (QFIM), which is essential for describing the achievable precision in simultaneous multiparameter measurements. The relevant numerical results and the measurement precision on oscillation parameters are also obtained.

    The remainder of this article is organized as follows. In Sec.~\ref{sec:QFI}, we briefly introduce the basic concepts in QET, including the classical Fisher information, the CR bound and their generalizations to quantum systems. The difference in the QFI resulting from the basis transformation on the quantum state is discussed, which is also illustrated explicitly with the system of two-flavor neutrino oscillations. In Sec.~\ref{sec:3flavor}, we compute the QFI for three-flavor neutrino oscillations in the flavor basis, providing both analytical expressions and numerical results, and further discuss the properties of the QFIM in multiparameter estimation, especially its impacts on the precision measurements of oscillation parameters. In particular, we examine the compatibility of the optimal measurements, the Holevo bound, and the spectral structure of the QFIM, which determine which parameter combinations can be estimated simultaneously with finite precision. In Sec.~\ref{sec:sum}, the main conclusions of the article are summarized.

	\section{QFI and neutrino oscillations}
	\label{sec:QFI}
	
	\subsection{Concepts of QFI}
	
	The concept of the QFI stems from the classical Fisher information. 
	The classical Fisher information describes the information that the observable random variable $x$ carries about the parameter $\lambda$ with a distribution of $x$ in specific measurements. 
	Formally, it is defined as~\cite{ParticleDataGroup:2026aaa}
	\begin{eqnarray}
		F_\lambda^{} \equiv \int p(x|\lambda) \left[\frac{\partial \ln p(x|\lambda)}{\partial\lambda}\right]^2 {\rm d}x = \int \frac{\left[\partial_\lambda^{} p(x|\lambda)\right]^2}{p(x|\lambda)} \,{\rm d}x \;,
	\end{eqnarray}
	where $p(x|\lambda)$ is the conditional probability of obtaining the value $x$ when the parameter is fixed at the value $\lambda$. 
	In quantum mechanics, for a specific quantum state $|\psi\rangle$ and its corresponding density matrix $\rho_\lambda^{} \equiv |\psi\rangle \langle \psi |$, the probability distribution of the measurements can be calculated as $p(x|\lambda)={\rm Tr}[\rho_\lambda^{} M_x^{}]$. 
	Here, the positive operator-valued measure $M_x^{}$ satisfies $\int {\rm d}x \, M_x^{}={\bf 1}$, where the measurement strategy $M_x^{}$ is fixed and the integration is performed over all possible outcomes $x$.
	Now, the classical Fisher information {\it for a fixed} $M_x^{}$ can be rewritten as
	\begin{eqnarray}
		\label{eq:CFI}
		F [M_x^{}] (\lambda) = \sum_x \frac{\left( {\rm Tr} \left[\left( \partial_\lambda \rho_\lambda^{}\right) M_x^{}\right] \right)^2}{{\rm Tr} \left[ \rho_\lambda^{} M_x^{}\right]} \;,
	\end{eqnarray}
	providing the classical CR bound on the precision of measurements.
	
	Based on Eq.~(\ref{eq:CFI}), the QFI is defined as the maximum value of the classical Fisher information among all possible strategies of measurements, i.e.,
	\begin{eqnarray}
        \label{eq:CFI_Qlambda}
		Q_\lambda^{} \equiv \max_{M_x^{}} F[M_x^{}](\lambda) \;.
	\end{eqnarray}
	Mathematically, the QFI quantifies the distinguishability between nearby quantum states that depend on a parameter of interest.
	For a system described by a density matrix $\rho_\lambda^{}$, the QFI $Q_\lambda^{}$ measures how rapidly the state changes with respect to the parameter $\lambda$.
	A larger QFI implies that even small variations in the parameter could result in more distinguishable states, thereby enabling a more precise estimation.
	Using Eqs.~(\ref{eq:CFI}) and (\ref{eq:CFI_Qlambda}), one may derive the upper bound of the classical Fisher information, i.e., the expressions for the QFI as\footnote{Detailed derivations can be found in, e.g., Refs.~\cite{Yuen:1973mjw,Braunstein:1994zz,Paris:2008zgg,Nogueira:2016qsk}.}
	\begin{eqnarray}
		Q_\lambda^{} = {\rm Tr} \left[\rho_\lambda^{} L_\lambda^2 \right] \;,
	\end{eqnarray}
	which is independent of the measurement strategy but depends only on the quantum state $|\psi\rangle$ and the density matrix $\rho_\lambda^{}$ themselves. 
	
    As a Hermitian operator, the symmetric logarithmic derivative (SLD) $L_\lambda^{}$ is defined through the Lyapunov equation
	\begin{eqnarray}
		\label{eq:partial_rho}
		\frac{\partial \rho_\lambda^{}}{\partial \lambda} = \frac{1}{2} \left(\rho_\lambda^{} L_\lambda^{} + L_\lambda^{} \rho_\lambda^{} \right) \;.
	\end{eqnarray}
	The most general solution to this type of equation is~\cite{Paris:2008zgg}
	\begin{eqnarray}
		L_\lambda^{} = 2 \int_0^\infty {\rm d}t~{\rm e}^{-\rho_\lambda^{} t} \left(\frac{\partial \rho_\lambda^{}}{\partial \lambda}\right) {\rm e}^{-\rho_\lambda^{} t} \;,
	\end{eqnarray}
	and the QFI can also be written as $Q_\lambda^{} = {\rm Tr} [\dot{\rho}_\lambda^{} L_\lambda^{}]$ with $\dot{\rho}_\lambda^{} \equiv {\rm d}\rho_\lambda^{}/{\rm d}\lambda$.
	It is clear that the QFI is the variance of the SLD operator~\cite{Liu:2016can} with the definition in Eq.~(\ref{eq:partial_rho}) and ${\rm Tr}[\rho_\lambda^{}]=1$.
	
	For a pure state with the density matrix $\rho_\lambda^{}=|\psi\rangle \langle \psi |$, we have $\rho_\lambda^2=\rho_\lambda^{}$ and 
	\begin{eqnarray}
		\dot{\rho}_\lambda^{} = \rho_\lambda^{} \dot{\rho}_\lambda^{} + \dot{\rho}_\lambda^{} \rho_\lambda^{} = \frac{1}{2} \left(\rho_\lambda^{} L_\lambda^{} + L_\lambda^{} \rho_\lambda^{} \right) \;,
	\end{eqnarray}
	from which $L_\lambda^{} = 2\dot{\rho}_\lambda^{}$ is obtained. Using $|\dot{\psi}\rangle \equiv |\partial_\lambda^{} \psi\rangle$, the QFI for pure states reads 
	\begin{eqnarray}
		\label{eq:QFI_pure_state}
		Q_\lambda^{} = 4 \left( \langle\dot{\psi} | \dot{\psi} \rangle - |\langle {\psi} | \dot{\psi} \rangle  |^2 \right) \;.
	\end{eqnarray}
	
	For multiparameter estimation, the definition of the QFI can be generalized to the QFIM~\cite{Liu:2019xfr,Albarelli:2020pec}.
	Assuming a density matrix $\rho_\lambda^{}$ depending on a set of parameters $\{\lambda_i^{}\}$, the matrix elements of the QFIM read~\cite{Helstrom:1967ldp}
	\begin{eqnarray}
		\left(Q_\lambda^{}\right)_{ij}^{} = {\rm Re}\;{\rm Tr}\left[\rho_\lambda^{} \frac{L_i^{} L_j^{} + L_j^{} L_i^{}}{2}\right] \;.
	\end{eqnarray}
	For a pure state $|\psi\rangle$, the corresponding elements of the QFIM are given by	
	\begin{eqnarray}
		\left(Q_\lambda^{}\right)_{ij}^{} \equiv 4\, {\rm Re} \left[\langle\partial_i^{} \psi | \partial_j^{} \psi \rangle - \langle \partial_i^{} \psi | \psi \rangle \langle \psi | \partial_j^{} \psi \rangle \right] \;.
	\end{eqnarray}
	It is obvious that off-diagonal elements satisfy the relation $Q_{ij}^{} = Q_{ji}^{}$, where the subscript referring to the parameter $\lambda$ has been suppressed and the dependence on $\lambda$ is implied, so that the QFIM is a real symmetric matrix.
	The off-diagonal elements indicate the overlap of parameter sensitivities in the quantum state. Furthermore, we denote the diagonal elements as $Q_i^{} \equiv Q_{ii}^{}$ for brevity in subsequent discussions.
	
	In the multiparameter case, we can also introduce the corresponding CR bound matrix, which describes the lower bound on the covariance between each two parameters
	\begin{eqnarray}
		{\rm Cov}[\lambda] \geqslant \frac{1}{N} Q_\lambda^{-1} \;,
	\end{eqnarray}
	with ${\rm Cov}[\lambda]_{ij}^{} \equiv \langle \lambda_i^{} \lambda_j^{} \rangle - \langle \lambda_i^{} \rangle \langle \lambda_j^{} \rangle$ and $\langle \lambda_i^{} \rangle$ being the expectation values of $\lambda_i^{}$.
	Diagonal elements ${\rm Cov}[\lambda]_{ii}^{}$ are lower bounds for the variances of $\lambda_i^{}$ with fixed values of other parameters, while the covariance matrix reduces exactly to the corresponding variance matrix if the QFIM is diagonal. 
	In general, since the quantum operators for measuring different parameters may be noncommutative, the CR bound cannot be achieved due to the quantum uncertainty relations.
	More precisely, the SLD--CR bound ${\rm Cov}[\lambda] \geqslant Q_\lambda^{-1}/N$ can be saturated asymptotically if and only if the SLD operators commute on average~\cite{Ragy:2016ycy,Albarelli:2020pec}, i.e.,
	\begin{eqnarray}
		\mathrm{Tr}\left[\rho_\lambda^{} [L_i^{}, L_j^{}]\right] = 0
	\end{eqnarray}
	for all parameter pairs $(i,j)$. For a pure state $|\psi\rangle$, with $L_i^{} = 2\left(|\partial_i^{}\psi\rangle\langle\psi| + |\psi\rangle\langle\partial_i^{}\psi|\right)$, this expectation value is proportional to the imaginary part of the quantum geometric tensor, namely
	\begin{eqnarray}
		\label{eq:commutator}
		\mathrm{Tr}\left[\rho_\lambda^{} [L_i^{}, L_j^{}]\right] = 8\,\mathrm{i}\,\mathrm{Im}\left[\langle\partial_i^{}\psi|\partial_j^{}\psi\rangle - \langle\partial_i^{}\psi|\psi\rangle\langle\psi|\partial_j^{}\psi\rangle\right] \;.
	\end{eqnarray}
	The combination in the square brackets is the quantum geometric tensor $Q^{\rm GT}_{ij}$, whose real part equals $\left(Q^{}_\lambda\right)_{ij}^{}/4$ and whose imaginary part, when nonvanishing, signals that the optimal measurements for $\lambda_i^{}$ and $\lambda_j^{}$ are incompatible, so that the SLD bound cannot be attained simultaneously for the two parameters.

    \subsection{QFI in different bases}
    
    In quantum mechanics, one often performs the transformation on a given quantum state expressed in one specific basis $|\psi (L)\rangle$ into another, i.e., $|\psi(L)\rangle \to {\cal U} |\psi(L)\rangle$ with ${\cal U}$ being an arbitrary unitary matrix. 
	It is well-known that such a unitary transformation applied simultaneously to both quantum states and observables does not change physical results, such as the expectation values. However, the value of the QFI may be changed as a consequence~\cite{Liu:2015zxo}. Take the QFI with respect to the parameter $\lambda$ as an example. If the unitary transformation depends on $\lambda$, then the derivative of the quantum state $|\dot{\psi}\rangle \equiv |\partial_\lambda^{} \psi\rangle$ changes as
	\begin{eqnarray}
		|\dot{\psi}\rangle \to \dot{{\cal U}}(\lambda) |{\psi}\rangle + {\cal U}(\lambda) |\dot{\psi}\rangle = {\cal U}(\lambda) \left[-{\rm i} {\cal R}(\lambda) |\psi\rangle + |\dot{\psi}\rangle \right] \;,
	\end{eqnarray}
	with the defined Hermitian operator ${\cal R}(\lambda) \equiv {\rm i}\,{\cal U}^\dagger(\lambda)\,\dot{{\cal U}}(\lambda)$ and $\dot{{\cal U}}(\lambda) \equiv \partial_\lambda^{} {\cal U}(\lambda)$. With the help of Eq.~(\ref{eq:QFI_pure_state}), we arrive at the QFI after the transformation
	\begin{eqnarray}
		\label{eq:QFI_after_U}
		Q_\lambda^{} &\to&  4 \left[\langle \dot{\psi} | \dot{\psi} \rangle - |\langle {\psi} | \dot{\psi} \rangle|^2 \nonumber \right. \\
		&& \left. + \langle \psi | {\cal R}^2 | \psi\rangle + \langle \psi | {\rm i} {\cal R} | \dot{\psi} \rangle - \langle \dot{\psi} | {\rm i} {\cal R} | \psi \rangle - \left( \langle \psi | {\cal R} | \psi\rangle \right)^2 - 2 \langle \psi | {\rm i} {\cal R} | {\psi} \rangle \langle \psi | \dot{\psi} \rangle \right] \;.
	\end{eqnarray}
	The terms in the first line represent the QFI computed in the original physical basis, whereas the ones in the second line highlight the discrepancy in the QFI values obtained in two different bases.
	On the other hand, if the unitary transformation ${\cal U}$, which connects two bases, is independent of the parameter $\lambda$, then the Hermitian matrix ${\cal R}={\bf 0}$, and in this case, the QFI does not change.
	Similarly, for the QFIM in the multiparameter case, the matrix elements after basis transformations read
	\begin{eqnarray}
		Q_{ij}^{} &=&  4\, {\rm Re} \left[\langle \dot{\psi} | {\psi}' \rangle - \langle {\psi} | \dot{\psi} \rangle \langle {\psi}' | {\psi} \rangle 
		+ \langle \psi | {\cal R}^\dagger {\cal G} | \psi\rangle + \langle \psi | {\rm i} {\cal R} | {\psi}' \rangle - \langle \dot{\psi} | {\rm i} {\cal G} | \psi \rangle \nonumber \right. \\
		&& \left. - \langle \psi | {\cal R} | \psi\rangle \langle \psi | {\cal G} | \psi\rangle  -  \langle \psi | {\rm i} {\cal R} | {\psi} \rangle \langle \psi | {\psi}' \rangle - \langle \psi | {\rm i} {\cal G} | {\psi} \rangle \langle \psi | \dot{\psi} \rangle \right] \;.
	\end{eqnarray}
	Here, we define $|\dot{\psi}\rangle \equiv |\partial_i^{} \psi\rangle$ and $|\psi'\rangle \equiv |\partial_j^{} \psi\rangle$ with ${\cal R} \equiv {\rm i}\,{\cal U}^\dagger \dot{{\cal U}}$ and ${\cal G} \equiv {\rm i}\,{\cal U}^\dagger {{\cal U}}'$. Choosing the real part of the expression guarantees the symmetric property of the QFIM.
	
	Although, according to Eq.~(\ref{eq:QFI_after_U}), performing basis transformation on the physical states does change the QFI, this is not always the case.
	A simple example is when ${\cal U}(\lambda) = {\rm e}^{-{\rm i} \lambda} {\bf 1}$, i.e., only an overall phase redefinition of the quantum state is performed.	In this case, even though the parameter $\lambda$ resides in the unitary transformation ${\cal U}(\lambda)$, the Hermitian matrix ${\cal R}={\bf 1}$, and it is evident that the QFI remains unchanged as long as $\langle \psi | \psi \rangle = 1$ holds.
	Once the dependence of ${\cal U}(\lambda)$ on $\lambda$ becomes more intricate, the QFI evaluated in different physical bases will no longer coincide. We will demonstrate this with the example of two-flavor neutrino oscillations.
	
	\subsection{QFI for two-flavor neutrino oscillations}
	
	First, we start our discussion in the neutrino mass basis. In this case, two orthogonal mass eigenstates can be fixed as
	\begin{eqnarray}
		|\nu_1^{}\rangle = \begin{pmatrix}
			1 \\ 0
		\end{pmatrix} \;, \quad |\nu_2^{}\rangle = \begin{pmatrix}
		0 \\ 1
		\end{pmatrix} \;.
	\end{eqnarray}
	The effective Hamiltonian and the two-flavor mixing matrix are accordingly given by
	\begin{eqnarray}
		H_0^{} = \frac{1}{2E} \begin{pmatrix}
			m_1^2 & 0 \\ 0 & m_2^2
		\end{pmatrix} \;, \quad U=\begin{pmatrix}
		\cos\theta & \sin\theta \\ -\sin\theta & \cos\theta
		\end{pmatrix} \;,
	\end{eqnarray}
	with $\{m^{}_{1}, m^{}_2\}$ being two neutrino mass eigenvalues, the neutrino energy $E$ and only one mixing angle $\theta$.
	Thus, the time-evolved electron neutrino state in the mass basis $|\psi_{\nu_e^{}}^{} (L)\rangle_{\rm m}$ with the baseline length $L$ can be expressed (up to an overall phase factor) as
	\begin{eqnarray}
		|\psi_{\nu_e^{}}^{} (L)\rangle_{\rm m} = \cos\theta |\nu_1^{}\rangle + \sin\theta \, {\rm e}^{-{\rm i} \Delta_{21}^{}} |\nu_2^{} \rangle = \begin{pmatrix}
			\cos\theta \\ \sin\theta \,{\rm e}^{-{\rm i} \Delta_{21}^{}}
		\end{pmatrix} \;,
	\end{eqnarray}
	with $\Delta_{21}^{} \equiv \Delta m_{21}^2 L / (2E)$ and $\Delta m^2_{21} \equiv m^2_2 - m^2_1$. With the help of Eq.~(\ref{eq:QFI_pure_state}), we arrive at the QFI on the mixing angle $\theta$, the mass-squared difference $\Delta m_{21}^2$, and the off-diagonal element, respectively:
	\begin{eqnarray}
		Q_{\theta,{\rm m}}^{\nu_e^{}} = 4 \;, \quad Q_{\Delta m_{21}^2,{\rm m}}^{\nu_e^{}} = \left(\frac{L }{E}\right)^2 \sin^2 \theta \cos^2 \theta  \;, \quad Q_{\Delta m_{21}^2 \theta,{\rm m}}^{\nu_e^{}} = 0 \;.
	\end{eqnarray}
	Notice that the neutrino state and the QFI in the mass basis will be labeled by the subscript ``m", while those in the flavor basis by the subscript ``f\ ". The QFI for the muon neutrino beam can be calculated in the same way, which are the same as those for the electron neutrino beam. 
	
	Then, we turn to the flavor basis, in which the orthogonal eigenstates are
	\begin{eqnarray}
		|\nu_e^{}\rangle = \begin{pmatrix}
			1 \\ 0
		\end{pmatrix} \;, \quad |\nu_\mu^{}\rangle = \begin{pmatrix}
			0 \\ 1
		\end{pmatrix} \;.
	\end{eqnarray}
	Now, the effective Hamiltonian in the flavor basis is described by
	\begin{eqnarray}
		H = U H_0^{} U^\dagger = \frac{\Delta m_{21}^2}{4E} \begin{pmatrix}
			-\cos2\theta & \sin2\theta \\ \sin2\theta & \cos2\theta
		\end{pmatrix} \;,
	\end{eqnarray}
	after subtracting the irrelevant terms proportional to the identity matrix. As the electron neutrino state propagates a distance $L$, it becomes
	\begin{eqnarray}
		|\psi_{\nu_e^{}}^{} (L) \rangle_{\rm f} = {\rm e}^{-{\rm i} H L} |\nu_e^{}\rangle = \begin{pmatrix}
			\cos^2\theta + {\rm e}^{- {\rm i} \Delta_{21}^{}} \sin^2\theta \\
			\sin\theta \cos\theta \left({\rm e}^{- {\rm i} \Delta_{21}^{}} - 1 \right) 
		\end{pmatrix} \;,
	\end{eqnarray}
	where we have used the relation ${\rm e}^{-{\rm i} H L} = U {\rm e}^{-{\rm i} H_0^{} L} U^\dagger$. 
	The diagonal and off-diagonal elements of the QFIM in the flavor basis can now be calculated in a similar way
	\begin{eqnarray}
		Q_{\theta,{\rm f}}^{\nu_e^{}} &=& 16 \sin^2\left(\frac{\Delta m_{21}^2 L}{4 E}\right) \left[1-\sin^2 2 \theta \cos^2\left(\frac{\Delta m_{21}^2 L}{4 E}\right)\right] \;, \nonumber \\
		Q_{\Delta m_{21}^2,{\rm f}}^{\nu_e^{}} &=& \left(\frac{L }{E}\right)^2 \sin^2 \theta \cos^2 \theta  \;, \quad Q_{\Delta m_{21}^2 \theta,{\rm f}}^{\nu_e^{}} = \frac{L}{2 E} \sin 4 \theta \sin\left(\frac{\Delta m_{21}^2 L}{2 E}\right) \;.
	\end{eqnarray}
	For the QFI of muon neutrinos, the quantum state $|\psi_{\nu_\mu^{}}^{} (L)\rangle_{\rm f}$ can be obtained by simply replacing $\theta \to -\theta$ and $\Delta m_{21}^2 \to -\Delta m_{21}^2$ in $|\psi_{\nu_e^{}}^{} (L)\rangle_{\rm f}$, then it is easy to see that all results are the same as those of electron neutrinos.
	
	Comparing the results calculated in the mass and flavor bases, we find that the QFI on $\Delta m_{21}^2$ remains the same, since the PMNS matrix $U$ does not depend on $\Delta m_{21}^2$, while that for $\theta$ and the off-diagonal matrix elements differ from each other. More specifically, the QFI on $\theta$ in the mass basis $Q_{\theta,{\rm m}}^{\nu_e^{}}$ is a constant number, independent of the oscillation length $L$, the neutrino energy $E$ and the mixing angle $\theta$, while the result in the flavor basis $Q_{\theta,{\rm f}}^{\nu_e^{}}$ oscillates with respect to $L/E$. This difference can be described by the aforementioned fact that performing a unitary transformation ${\cal U}$ on the quantum state $|\psi\rangle$, which depends on the parameter of our interest, could modify the QFI. If we convert the state $|\psi_{\nu_e^{}}^{} (L) \rangle_{\rm m}$ in the mass basis to $|\psi_{\nu_e^{}}^{} (L) \rangle_{\rm f}$ in the flavor basis with the unitary matrix ${\cal U} = U$, then the Hermitian matrix ${\cal R}$ is just the second Pauli matrix $\sigma^{}_2$. Using Eq.~(\ref{eq:QFI_after_U}), it is straightforward to prove that the QFI changes from $Q_{\rm m}^{\nu_e^{}}$ to $Q_{\rm f}^{\nu_e^{}}$.
   	
	\section{Three-flavor neutrino oscillations}
	\label{sec:3flavor}
	
	\begin{table}[t]
		\centering
		\begin{tabular}{cccccc}
			\toprule
			& DUNE & T2HK & ESSnuSB & JUNO & Daya Bay \\ \midrule
			$L~[{\rm km}]$   & 1300 & 295 & 540 & 53 & 1.7 \\
			$E~[{\rm GeV}]$   & 2.5     &  0.6   &   0.36      &  4$\times 10^{-3}$     &   4$\times 10^{-3}$       \\
			$L/E~[{\rm km}/{\rm GeV}]$ &  520    &  492   &   1500      &  13250    &   425       \\ \bottomrule
		\end{tabular}
		\caption{The baseline length $L$ and the peak energy $E$ for various neutrino oscillation experiments.}
		\label{tab:L/E}
	\end{table}
	
	Next, we consider the case of three-flavor neutrino oscillations, in which the flavor mixing of massive neutrinos is described by the $3\times 3$ PMNS matrix in Eq.~(\ref{eq:PMNS}).
	We choose to calculate the QFI in the neutrino flavor basis, as the detection of neutrinos is sensitive to different flavors.
	We also apply the QFI to existing and future neutrino oscillation experiments. For this purpose, the baseline lengths and the peak energies corresponding to typical experiments are listed in Table~\ref{tab:L/E}. Throughout this work the neutrino oscillation amplitudes are evaluated in vacuum, i.e., matter effects are neglected. The results associated with the various experiments should therefore be understood as state-level quantum-information benchmarks for these configurations, rather than as realistic experimental sensitivity projections, which would additionally require a full treatment of matter effects and of the detector response.
	
	\subsection{Electron neutrinos}
	
	First, we discuss the QFI for electron neutrinos. In the case of three neutrino flavors, there are six oscillation parameters, i.e., three mixing angles $\{\theta^{}_{12}, \theta^{}_{13}, \theta^{}_{23}\}$, two mass-squared differences $\{\Delta m^2_{21}, \Delta m^2_{31}\}$, and one {\it CP}-violating phase $\delta^{}_{\rm CP}$. Therefore, the QFIM is a $6 \times 6$ matrix. The quantum state for an initial electron neutrino after propagating a distance $L$ is
	\begin{eqnarray}
		|\psi_{\nu_e^{}}^{} (L) \rangle_{\rm f}^{} = \begin{pmatrix}
			s_{12}^2 c_{13}^2 {\rm e}^{-{\rm i} \Delta_{21}^{}}+s_{13}^2 {\rm e}^{-{\rm i} \Delta_{31}^{}}+c_{12}^2 c_{13}^2 \vspace{0.15cm} \\
			c_{13}^{} \left[ s_{12}^{} c_{12}^{} c_{23}^{} \left({\rm e}^{- {\rm i} \Delta_{21}^{}}-1\right) - {\rm e}^{{\rm i} \delta_{\rm CP}^{} } c^2_{12} s_{13}^{} s_{23}^{} + {\rm e}^{{\rm i} \delta_{\rm CP}^{} } s_{13}^{} s_{23}^{} \left({\rm e}^{- {\rm i} \Delta_{31}^{}}-s^2_{12} {\rm e}^{- {\rm i} \Delta_{21}^{}}\right)\right] \vspace{0.15cm} \\
			c_{13}^{} \left[{\rm e}^{{\rm i} \delta_{\rm CP}^{} } s_{13}^{} c_{23}^{} \left({\rm e}^{-{\rm i} \Delta_{31}^{}} - s_{12}^2 {\rm e}^{-{\rm i} \Delta_{21}^{}} - c_{12}^2\right) + s_{12}^{} c_{12}^{} s_{23}^{} \left(1-{\rm e}^{-{\rm i} \Delta_{21}^{}}\right)\right]
		\end{pmatrix} \;, \qquad
	\end{eqnarray}
	with $\Delta_{ji}^{} \equiv \Delta m_{ji}^2 L/(2E)$ for $j,i=2,1;3,1$. Based on such a time-evolved quantum state, we can calculate all diagonal and off-diagonal elements of the QFIM with the help of Eq.~(\ref{eq:QFI_pure_state}).
	For example, the diagonal QFI on $\theta_{13}^{}$ is given by
	\begin{eqnarray}
		\label{eq:Qtheta13nue}
		Q_{\theta_{13}^{}}^{\nu_e^{}}  &\simeq &  16 \sin ^2\left(\frac{\Delta m_{31}^2 L}{4 E}\right) - 4   s^2_{12} \frac{ \Delta m_{21}^2 L}{E} \sin \left(\frac{\Delta m_{31}^2 L}{2 E}\right) -16 s_{13}^2 \sin^2\left(\frac{\Delta m_{31}^2 L}{2E}\right)   \nonumber \\
		&& + \frac{64 s_{13}^4}{3} \sin ^2\left(\frac{\Delta m_{31}^2 L}{2 E}\right) + s_{12}^2 \left(\frac{\Delta m_{21}^2 L}{E}\right)^2  \left[\cos \left(\frac{\Delta m_{31}^2 L}{2 E}\right) - c_{12}^2 \right] \nonumber \\
		&& + 8 s_{12}^2 s_{13}^2 \left(\frac{\Delta m_{21}^2 L}{E}\right) \sin \left(\frac{\Delta m_{31}^2 L}{E}\right) \;,
	\end{eqnarray}
	via a series expansion in the small parameters $\sin^2\theta_{13}^{}$ and $\Delta m_{21}^2/\Delta m_{31}^2$, using the method proposed in Ref.~\cite{Akhmedov:2004ny}. Especially, we present the series expansions up to {\it first} or {\it second} order, depending on how well the behavior of the exact numerical results can be described.
	
	\begin{figure}[t]
		\centering
		\includegraphics[scale=0.45]{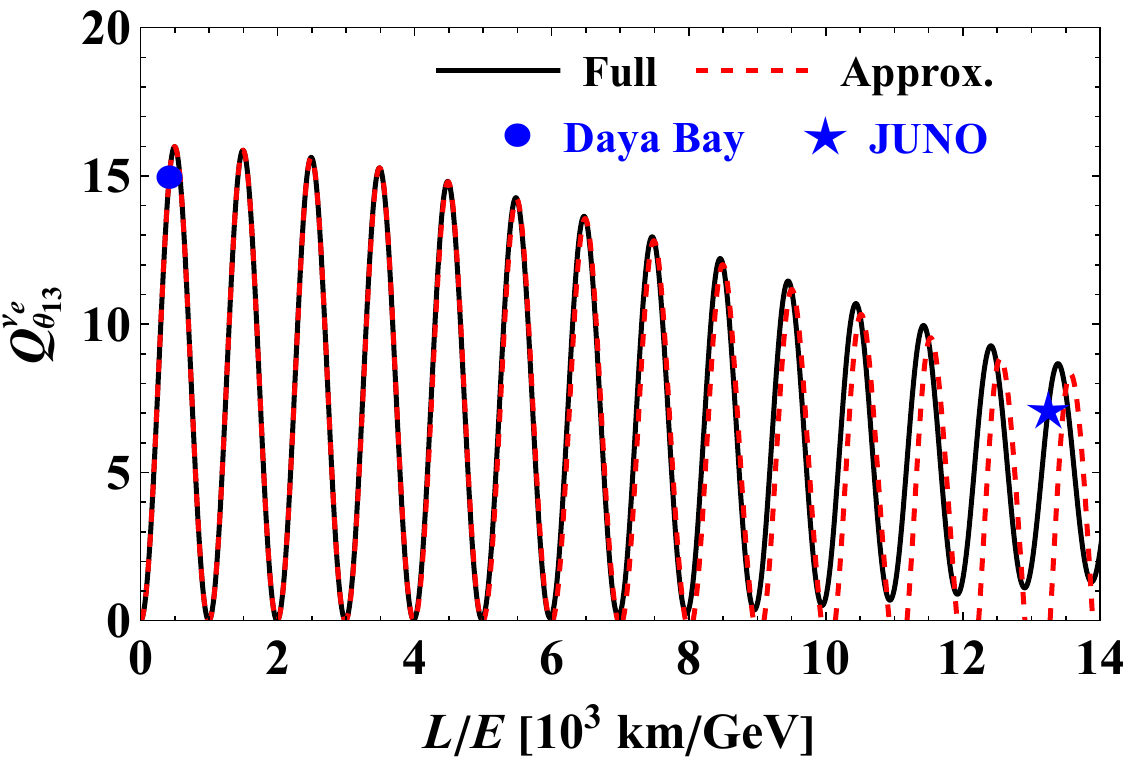}
		\includegraphics[scale=0.45]{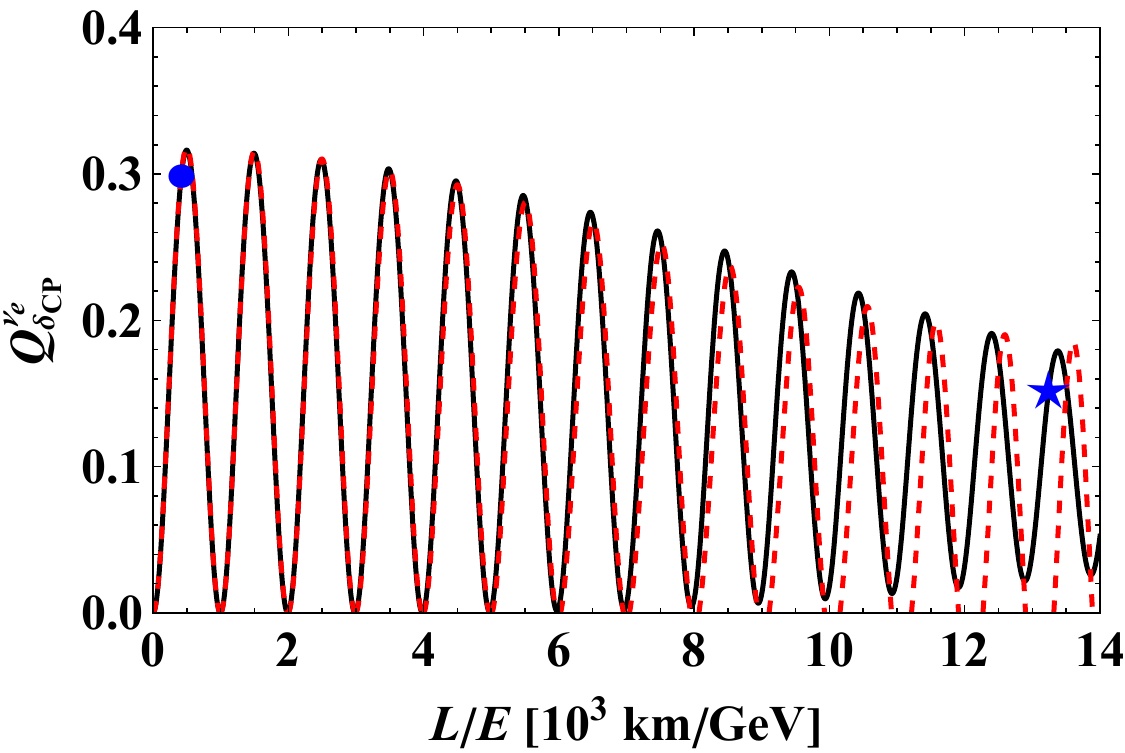}
		\vspace{-0.1cm}
		\caption{The numerical values of $Q_{\theta_{13}^{}}^{\nu_e^{}}$ (left panel) and $Q_{\delta_{\rm CP}^{}}^{\nu_e^{}}$ (right panel) as functions of $L/E$. 
		The full numerical values are plotted as black solid curves, while the approximate results from the series expansion in both $\sin^2\theta_{13}^{}$ and $\Delta m_{21}^2/\Delta m_{31}^2$ are plotted as red dashed curves. The blue dot and the blue star correspond to the QFI values computed with the $L/E$ associated with Daya Bay and JUNO, respectively.}
		\label{fig:hnue}
	\end{figure}
	
	In the left panel of Fig.~\ref{fig:hnue}, the numerical results for $Q_{\theta_{13}^{}}^{\nu_e^{}}$ using the full expression and the approximate one in Eq.~(\ref{eq:Qtheta13nue}) are plotted as the black solid curve and the red dashed curve, respectively. The numerical inputs for the oscillation parameters are chosen from the latest global analyses of neutrino oscillation experiments~\cite{Esteban:2024eli,Esteban:2026phq,NuFIT:6-1}. Specifically, we consider the results including the Super-Kamiokande atmospheric neutrino data in the NO case~\cite{NuFIT:6-1}, namely, 
	\begin{eqnarray}
		\left\{\theta_{12}^{}, \theta_{13}^{}, \theta_{23}^{}, \delta_{\rm CP}^{}\right\} &=& \left\{33.76^\circ, 8.62^\circ, 43.29^\circ, 212^\circ\right\} \;, \nonumber \\
		\left\{\Delta m_{21}^2, \Delta m_{31}^2\right\} &=& \left\{7.537 \times 10^{-5}~{\rm eV}^2, 2.511 \times 10^{-3}~{\rm eV}^2\right\} \;.
	\end{eqnarray}
	It can be observed from Fig.~\ref{fig:hnue} that our series expansion in $\sin^2\theta_{13}^{}\approx 0.022$ and $\Delta m_{21}^2/\Delta m_{31}^2 \approx 0.03$ successfully captures the oscillation behavior of the QFI for small $L/E$.
	At relatively large $L/E$, there is a visible discrepancy between the analytic approximation and the numerical result, since the terms proportional to $L/E$ amplify the deviations induced by the series expansion. Those deviations could be ascribed to higher-order contributions. Moreover, the amplitude gradually decreases at large $L/E$ due to the negative term proportional to $-s_{12}^2 c_{12}^2 (L/E)^2$ in the second line of Eq.~(\ref{eq:Qtheta13nue}).
	
	Thus, this calculation could help provide the information about $\theta_{13}^{}$ extracted from reactor neutrino experiments and the precision of its measurement. In particular, we also indicate the QFI values for two reactor neutrino experiments, i.e., Daya Bay and JUNO, as the blue dot and the blue star, respectively. Notice that the QFI for electron antineutrinos can be obtained by replacing $\delta^{}_{\rm CP} \to -\delta^{}_{\rm CP}$. Since Eq.~(\ref{eq:Qtheta13nue}) is independent of $\delta^{}_{\rm CP}$, it applies to electron antineutrinos as well and the difference appears only at even higher orders. For the Daya Bay experiment with $L/E \approx 425~{\rm km}/{\rm GeV}$ and approximately five million events~\cite{DayaBay:2025ngb}, the CR bound yields a lower limit on the experimental error of $\sigma [\theta_{13}^{}] \equiv \sqrt{{\rm Var}[\theta_{13}^{}]} \gtrsim 0.006^\circ $.
	
	Meanwhile, we calculate the QFI on $\delta_{\rm CP}^{}$ as
	\begin{eqnarray}
		Q_{\delta_{\rm CP}^{}}^{\nu_e^{}} &\simeq & \frac{s_{13}^2 }{12} \left\{32 \left(3-13 s_{13}^2\right)+4 \cos \left(\frac{\Delta m_{31}^2 L}{2 E}\right) \left[8 \left(16 s_{13}^2-3\right) + 3 s_{12}^2 \left(\frac{\Delta m_{21}^2 L}{E}\right)^2 \right] \right. \nonumber \\
		&& \left. -48 \left[s_{12}^2  \frac{\Delta m_{21}^2 L}{E}  \sin \left(\frac{\Delta m_{31}^2 L}{2 E}\right)+2  s_{13}^2 \cos \left(\frac{\Delta m_{31}^2 L}{E}\right)\right] - 3 \sin^2 2 \theta_{12}^{} \left(\frac{\Delta m_{21}^2 L}{E}\right)^2 \right\} \;. \qquad
	\end{eqnarray}
	The numerical values of $Q_{\delta_{\rm CP}^{}}^{\nu_e^{}}$ are plotted in the right panel of Fig.~\ref{fig:hnue} with the same legends as in the left panel. In this case, we series expand the analytical formula up to {\it second} order in $\sin^2\theta_{13}^{}$ and $\Delta m_{21}^2/\Delta m_{31}^2$, so that the approximate result is well consistent with the numerical one. Moreover, it is evident that the QFI on $\delta_{\rm CP}^{}$ is several orders of magnitude smaller than that on $\theta_{13}^{}$, indicating that more information about $\theta_{13}^{}$ can be extracted from the $\nu_e^{}$ beam. Therefore, the mixing angle $\theta_{13}^{}$ could be measured with high precision. However, from the perspective of QET, the precision in measuring $\delta_{\rm CP}^{}$ is inherently limited.
	
	It is worth noting that the QFI captures only the maximum information about the parameters in a neutrino beam, and the associated lower bound on measurement precision corresponds to the most ideal case. In practice, the sensitivity of a measurement is affected by many additional factors, which can result in an actual precision much below this bound or even a complete lack of sensitivity to the parameters. Even though the QFI for the electron neutrino beam on $\delta_{\rm CP}^{}$ is nonzero, and the values for Daya Bay and JUNO are close to the peak values, these experiments have no capability of measuring $\delta_{\rm CP}^{}$.
	
	\begin{figure}[t]
		\centering
   		\includegraphics[scale=0.45]{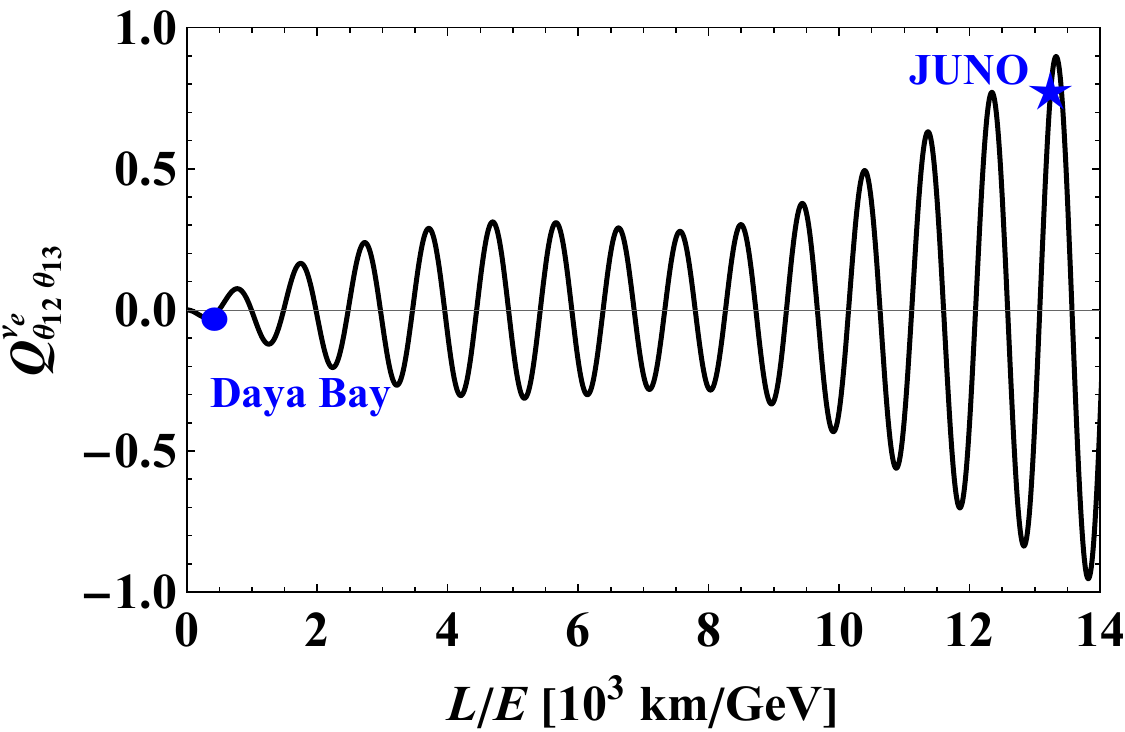}
		\includegraphics[scale=0.45]{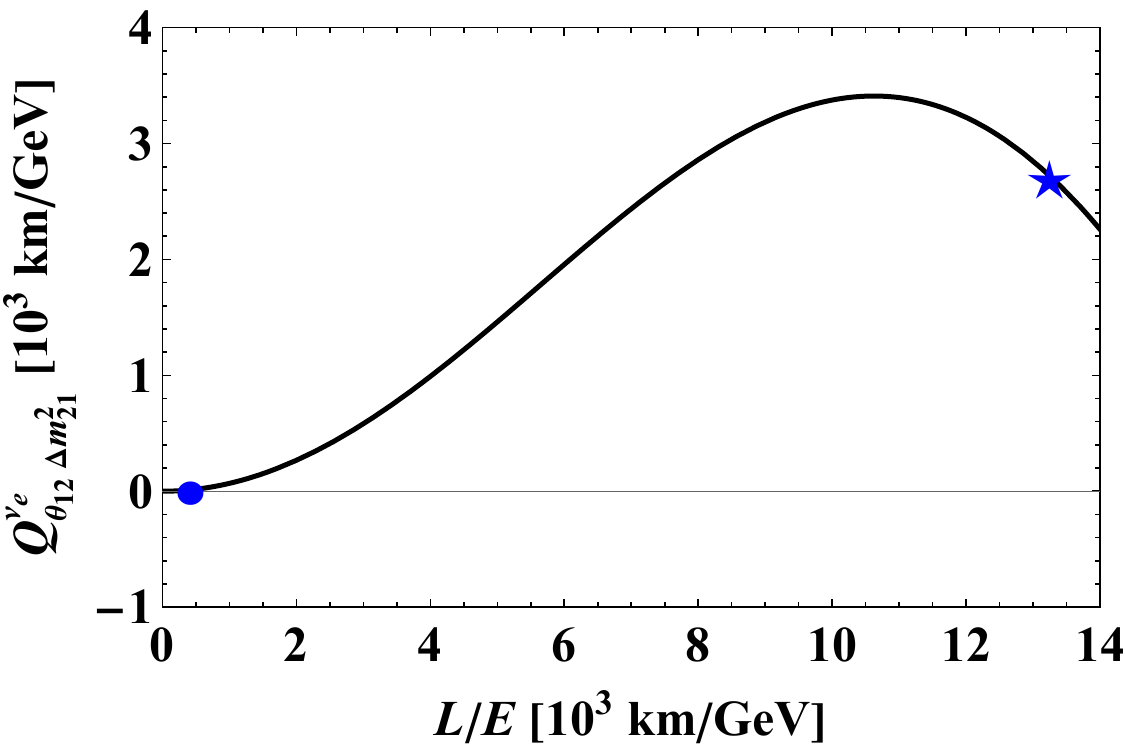}
		\vspace{-0.1cm}
		\caption{The off-diagonal elements $Q_{\theta_{12}^{} \theta_{13}^{}}^{\nu_e^{}}$ (left panel) and $Q_{\theta_{12}^{} \Delta m_{21}^2}^{\nu_e^{}}$ (right panel) of the QFIM as functions of $L/E$. The values of the QFI for Daya Bay and JUNO are shown as the blue dot and the blue star, respectively.}
		\label{fig:htheta12theta13nue}
	\end{figure}
   	
	Then, we calculate the off-diagonal elements of the QFIM. For example, the elements referring to $\theta_{12}^{}$-$\theta_{13}^{}$ and $\theta_{12}^{}$-$\Delta m_{21}^2$ for the electron neutrino beam are
	\begin{eqnarray}
		Q_{\theta_{12}^{} \theta_{13}^{}}^{\nu_e^{}} &= & - \sin 2 \theta_{12}^{} \sin 2 \theta_{13}^{} \sin \left(\frac{\Delta m_{21}^2 L}{4 E}\right) \left\{\left(\cos 2 \theta_{12}^{} - 2 c^2_{13}\right) \sin \left[\frac{(3 \Delta m_{21}^2 - 2 \Delta m_{31}^2) L}{4 E}\right] \right.  \nonumber \\
   		&& + (\cos 2 \theta_{12}^{}  + 2 c_{13}^{2}) \sin \left[\frac{ (\Delta m_{21}^2+2 \Delta m_{31}^2)L}{4 E}\right]+4 s_{13}^2 \sin \left[\frac{ (\Delta m_{21}^2-2 \Delta m_{31}^2)L}{4 E}\right]  \nonumber \\
   		&& \left. +2 \cos 2 \theta_{12}^{} \cos 2 \theta_{13}^{} \sin \left(\frac{\Delta m_{21}^2 L}{2 E}\right) \cos \left[\frac{ (\Delta m_{21}^2 - 2 \Delta m_{31}^2) L}{4 E}\right] \right\} \;,  \\
   		Q_{\theta_{12}^{} \Delta m_{21}^2}^{\nu_e^{}} &= &  \sin 2 \theta_{12}^{} c^2_{13} \left(c^2_{12}- s^2_{12} \cos 2 \theta_{13}^{} \right) \frac{L }{E} \sin \left(\frac{\Delta m_{21}^2 L}{2 E}\right) \;,
   	\end{eqnarray}
	which are plotted in the left and right panels of Fig.~\ref{fig:htheta12theta13nue}, respectively. It is worthwhile to stress that the exact analytical results are simple enough, so it is unnecessary to perform series expansions. For this reason, only the exact results are shown in Fig.~\ref{fig:htheta12theta13nue}, where the input parameters are the same as those in Fig.~\ref{fig:hnue}. It is evident that the off-diagonal elements of the QFIM are not always positive, which is not at all in conflict with its definition. In practice, the key quantity is the covariance matrix obtained by inverting the QFIM, as the signs of its off-diagonal elements reveal the statistical correlations between any two relevant parameters in question. This point will be explicitly demonstrated through numerical calculations in Sec.~\ref{subsec:QFIMandCR}.

    \subsection{Muon neutrinos}
	
	\begin{figure}[t]
		\centering
		\includegraphics[scale=0.36]{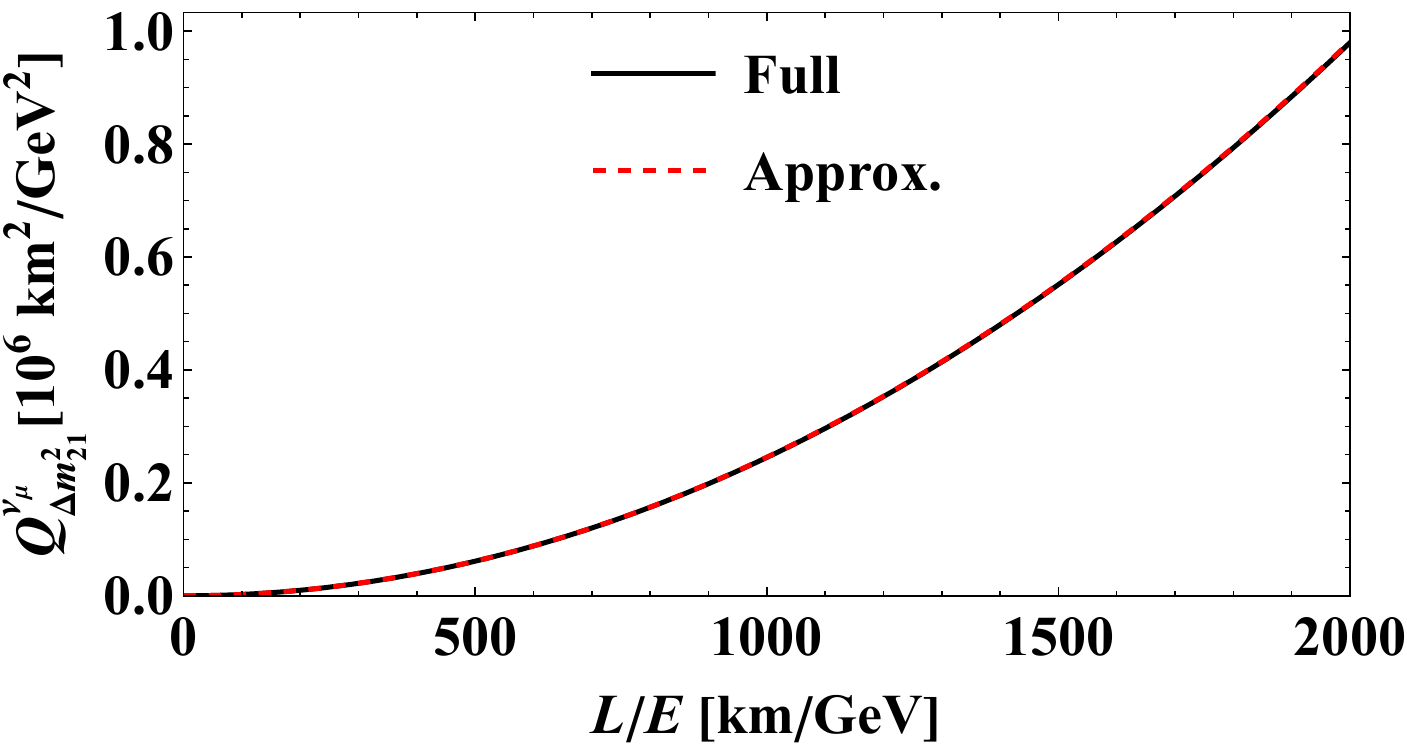}
		\includegraphics[scale=0.36]{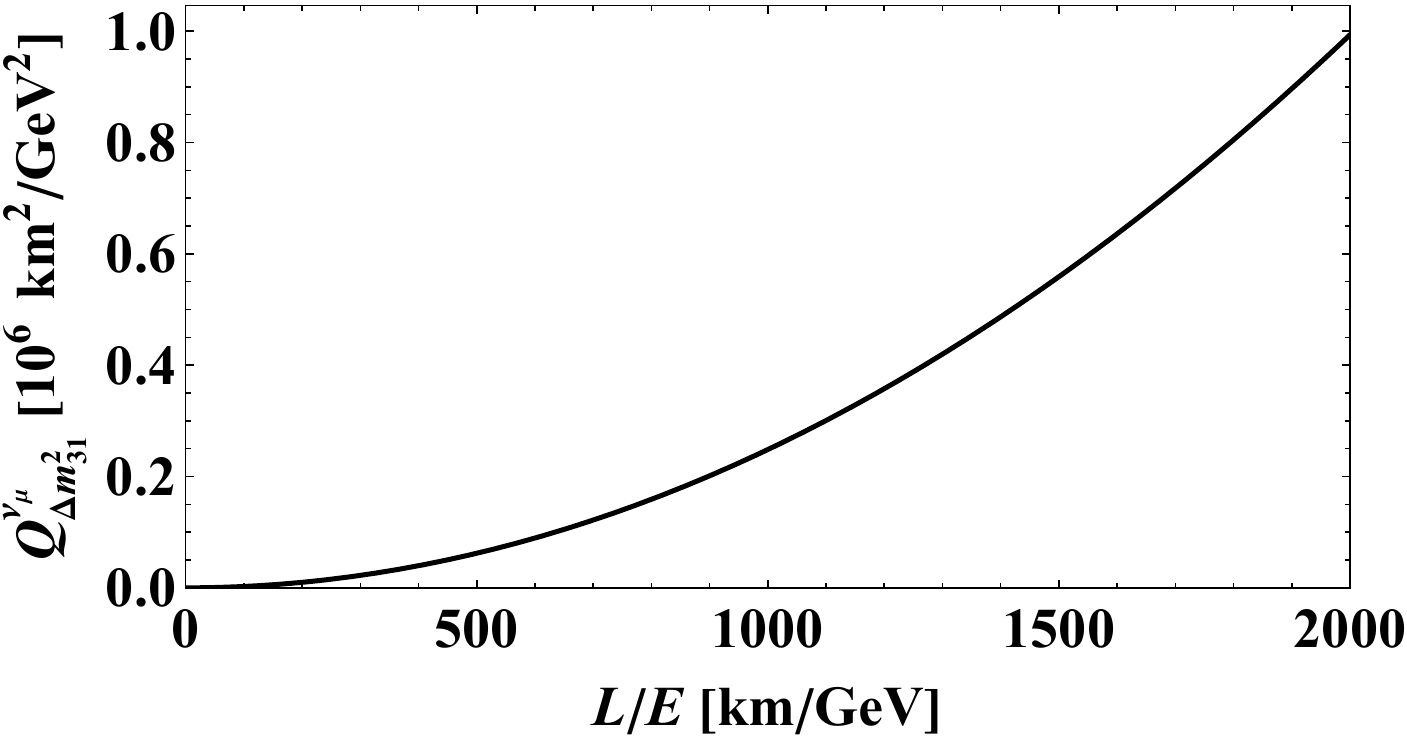} \\ 
		\includegraphics[scale=0.36]{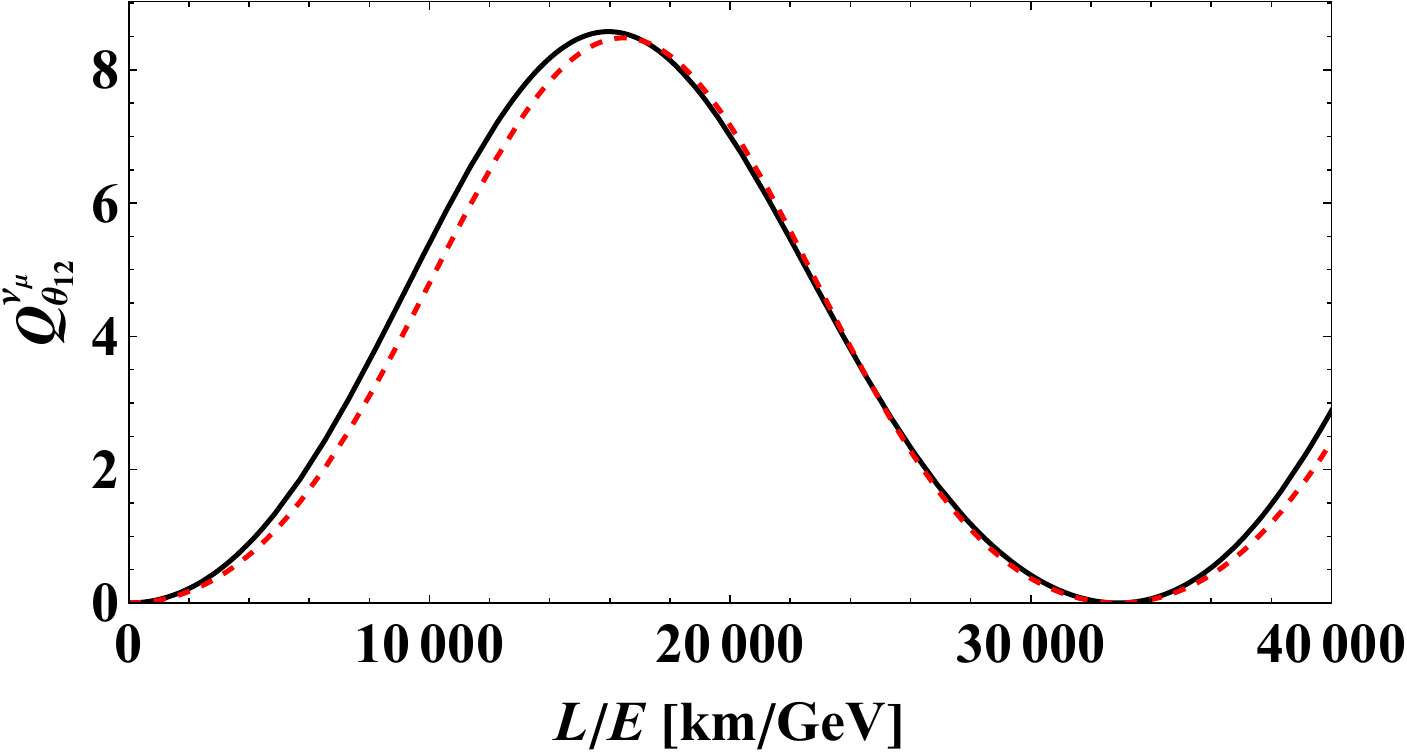}
		\includegraphics[scale=0.36]{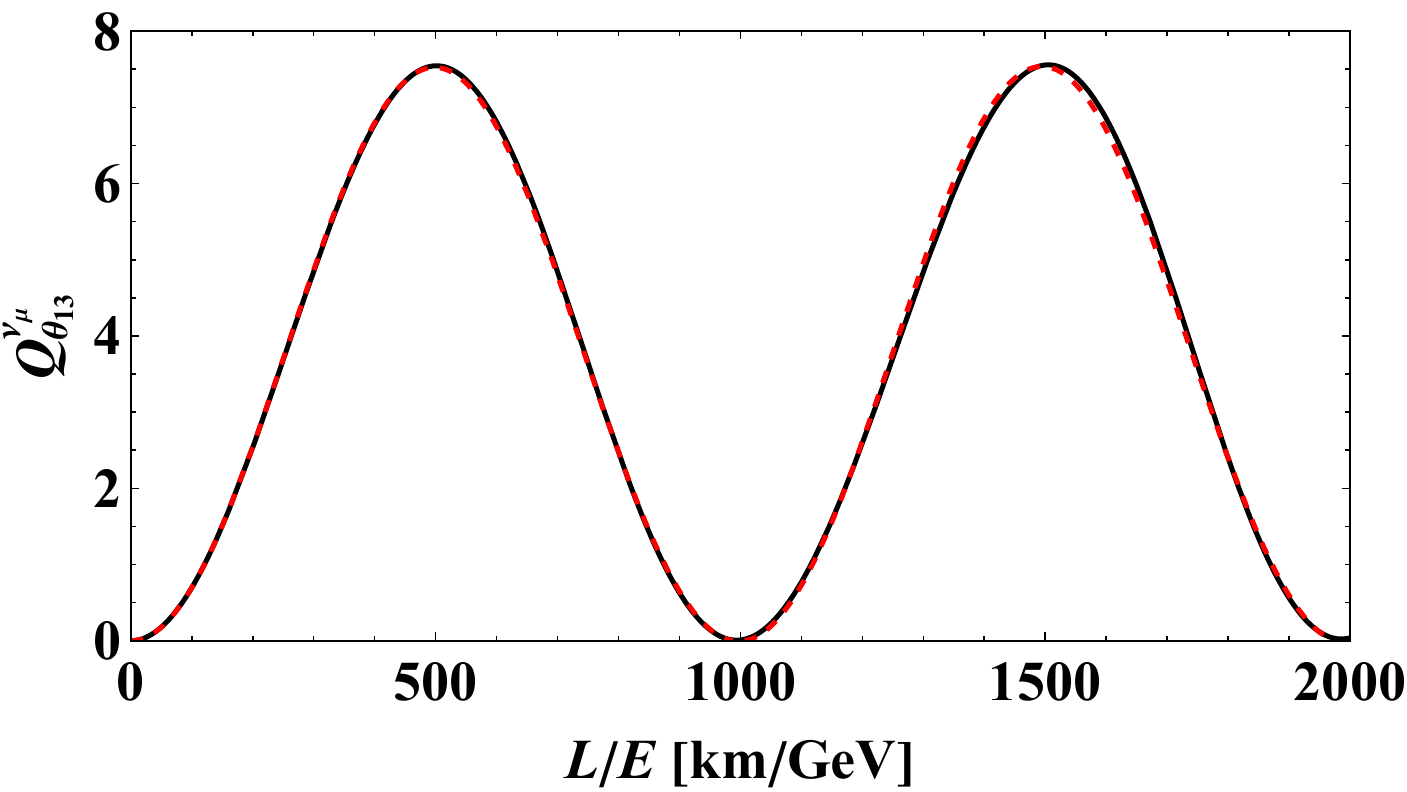}\\ \vspace{0.15cm}
		\includegraphics[scale=0.36]{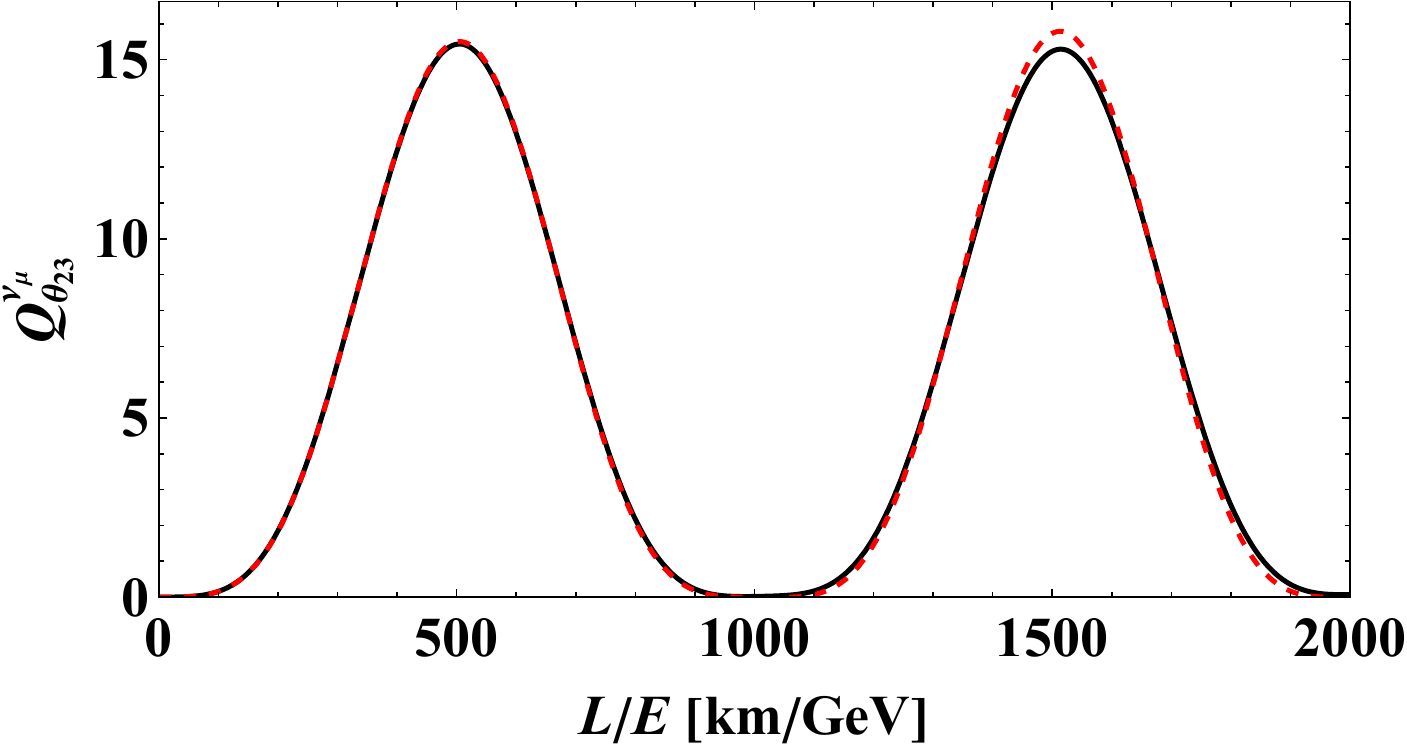}
		\includegraphics[scale=0.36]{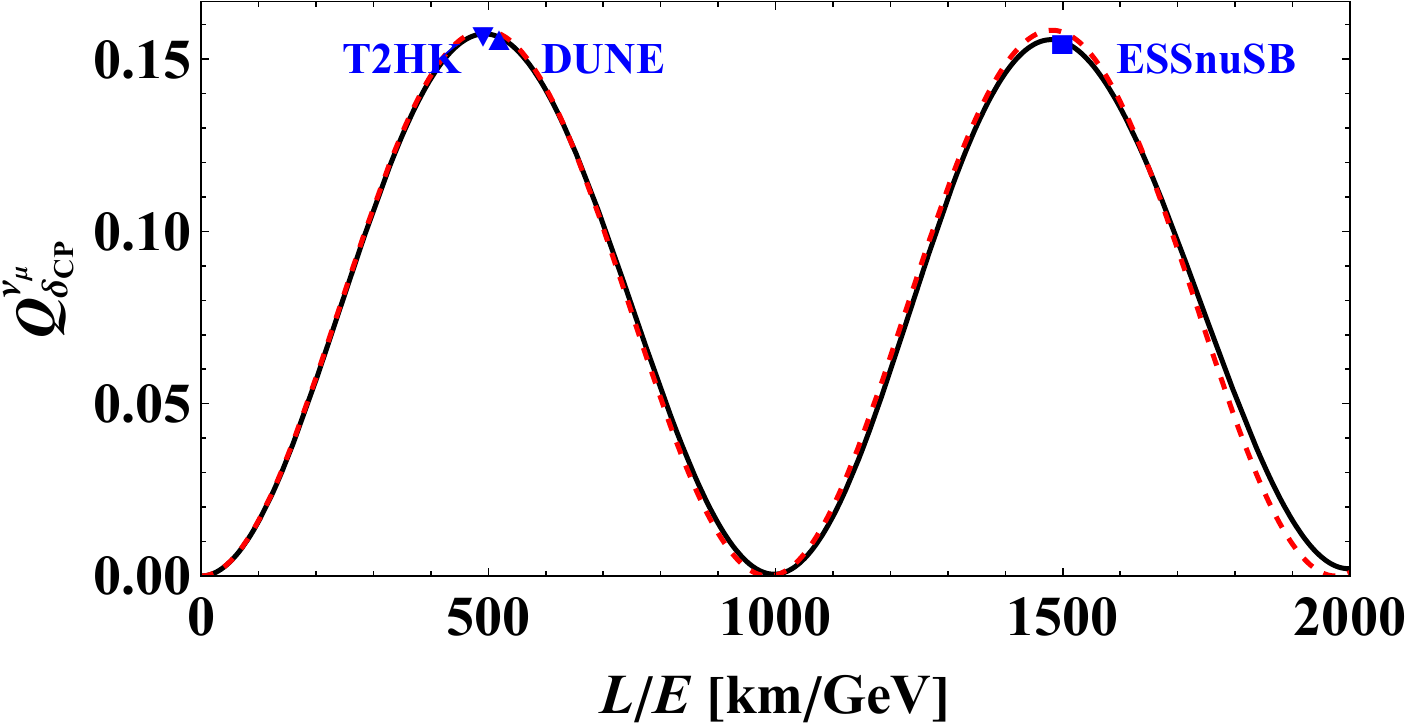}
		\vspace{-0.1cm}
		\caption{The QFI associated with different oscillation parameters for $\nu_\mu^{}$ with the same legends as those in Fig.~\ref{fig:hnue}.
			The range of $L/E$ in the plot of $Q_{\theta_{12}^{}}^{\nu_\mu^{}}$ is extended to $40\,000~{\rm km}/{\rm GeV}$ in order to fully capture its oscillation behavior.
            The values of $Q_{\delta_{\rm CP}^{}}^{\nu_\mu^{}}$ for DUNE, T2HK and ESSnuSB are labeled by the blue triangle, the blue inverted triangle and the blue square, respectively.}
		\label{fig:hnumu}
	\end{figure}
	
	Similarly to the analysis in the previous subsection, we consider the quantum state for an initial muon neutrino after propagating a distance $L$, which is given by
   	\begin{eqnarray}
   		|\psi_{\nu_\mu^{}}^{} (L) \rangle_{\rm f}^{} = \begin{pmatrix}
   			c_{13}^{} {\rm e}^{-{\rm i} \delta_{\rm CP}^{} } \left[ s_{13}^{} s_{23}^{}  \left({\rm e}^{- {\rm i} \Delta_{31}^{}} - s^2_{12} {\rm e}^{- {\rm i} \Delta_{21}^{}}\right) - c^2_{12} s_{13}^{} s_{23}^{}  + {\rm e}^{{\rm i} \delta_{\rm CP}^{} } s_{12}^{} c_{12}^{} c_{23}^{} \left({\rm e}^{- {\rm i} \Delta_{21}^{}} - 1 \right)\right] \vspace{0.15cm} \\ 
   			\begin{array}{l}
   				{\rm e}^{- {\rm i} \delta_{\rm CP}^{}} \left(c_{12}^{} s_{13}^{} s_{23}^{}+{\rm e}^{{\rm i} \delta_{\rm CP}^{} } s_{12}^{} c_{23}^{}\right) \left(s_{12}^{} c_{23}^{} +{\rm e}^{{\rm i} \delta_{\rm CP}^{} } c_{12}^{} s_{13}^{} s_{23}^{}\right) \\
   				+{\rm e}^{-{\rm i} \Delta_{21}^{}} \left(c_{12}^{} c_{23}^{}-{\rm e}^{-{\rm i} \delta_{\rm CP}^{} } s_{12}^{} s_{13}^{} s_{23}^{}\right) \left(c_{12}^{} c_{23}^{}-{\rm e}^{{\rm i} \delta_{\rm CP}^{} } s_{12}^{} s_{13}^{} s_{23}^{}\right) + c^2_{13} s^2_{23} {\rm e}^{-{\rm i} \Delta_{31}^{}} 
   			\end{array} \vspace{0.15cm} \\ 
   			\begin{array}{l}
   				\left(s_{12}^{} s_{23}^{}-{\rm e}^{{\rm i} \delta_{\rm CP}^{} } c_{12}^{} s_{13}^{} c_{23}^{}\right) \left(- s_{12}^{} c_{23}^{} - {\rm e}^{-{\rm i} \delta_{\rm CP}^{} } c_{12}^{} s_{13}^{} s_{23}^{}\right) + c^2_{13} s_{23}^{} c_{23}^{} {\rm e}^{-{\rm i} \Delta_{31}^{}} \\
   				+{\rm e}^{- {\rm i} \Delta_{21}^{}} \left(-c_{12}^{} s_{23}^{}-{\rm e}^{{\rm i} \delta_{\rm CP}^{} } s_{12}^{} s_{13}^{} c_{23}^{}\right) \left(c_{12}^{} c_{23}^{}-{\rm e}^{-{\rm i} \delta_{\rm CP}^{} } s_{12}^{} s_{13}^{} s_{23}^{}\right) 
   		\end{array}
   		\end{pmatrix} \quad
   	\end{eqnarray}
	with $\Delta_{ji}^{} \equiv \Delta m_{ji}^2 L/(2E)$ defined as before. 
   	Then, we have the QFI on the two mass-squared differences
   	\begin{eqnarray}
   		Q_{\Delta m_{31}^2}^{\nu_\mu^{}} &=&  c_{13}^2 s^2_{23} \frac{L^2 }{4 E^2} \left(2 c^2_{13} \cos 2 \theta_{23}^{} - \cos 2 \theta_{13}^{} + 3 \right) \;,  \\
   		Q_{\Delta m_{21}^2}^{\nu_\mu^{}} &\simeq& - c^2_{12} c^2_{23} \frac{L^2 }{4 E^2} \left(2 c^2_{12} \cos 2 \theta_{23}^{} + \cos 2 \theta_{12}^{} - 3\right) \nonumber \\
   		&& + s_{13}^{} \sin 2 \theta_{12}^{}  \sin 2 \theta_{23}^{} \cos\delta_{\rm CP}^{} \frac{ L^2 }{2 E^2} \left(c^2_{12} \cos 2 \theta_{23}^{} - s^2_{12 }\right) \nonumber \\
   		&& - s^2_{12} s^2_{13} s^2_{23} \frac{L^2  }{E^2} \left[2 c^2_{12} \left(c^2_{23} \cos 2\delta_{\rm CP}^{}  +\cos 2 \theta_{23}^{} \right)+\cos 2 \theta_{12}^{} \right] \;,
   	\end{eqnarray}
   	which are plotted in the first row of Fig.~\ref{fig:hnumu}.
   	The QFI on $\delta_{\rm CP}^{}$ reads
   	\begin{eqnarray}
   		Q_{\delta_{\rm CP}^{}}^{\nu_\mu^{}} &\simeq& -\frac{1}{8} \sin^2 2 \theta_{13}^{} \sin^2\left(\frac{\Delta m_{31}^2 L}{4 E}\right) \left[8 \sin ^2\left(\frac{\Delta m_{31}^2 L}{4 E}\right) \left(\cos 4 \theta_{13}^{} \cos 2 \theta_{23}^{} + 2 s^2_{13} c^2_{13} \cos 4 \theta_{23}^{} \right) \right. \nonumber \\
   		&& \left. -6 \cos 4 \theta_{13}^{} \sin ^2\left(\frac{\Delta m_{31}^2 L}{4 E}\right)
   		+ (4 \cos 2 \theta_{23}^{} - 3) \cos \left(\frac{\Delta m_{31}^2 L}{2 E}\right)+ 12 \cos 2 \theta_{23}^{} - 13 \right] \;, \qquad
   	\end{eqnarray}
   	while those on three mixing angles are
   	\begin{eqnarray}
   		Q_{\theta_{13}^{}}^{\nu_\mu^{}} &\simeq &  16 s^2_{23} \sin ^2\left(\frac{\Delta m_{31}^2 L}{4 E}\right) - 4 s_{12}^2 s_{23}^2 \frac{\Delta m_{21}^2 L }{E} \sin\left(\frac{\Delta m_{31}^2 L}{2 E}\right) - 16 s_{13}^2 s_{23}^4 \sin^2\left(\frac{\Delta m_{31}^2 L}{2E}\right) \;, \qquad \\
   		Q_{\theta_{12}^{}}^{\nu_\mu^{}} &\simeq & \frac{c^2_{23}}{2} \left\{2 c^2_{23} \left[2 \cos 4 \theta_{12}^{} \sin ^2\left(\frac{\Delta m_{21}^2 L}{2 E}\right)+\cos \left(\frac{\Delta m_{21}^2 L}{E}\right)\right] \right. \nonumber \\
   		&& \left. -16 \cos \left(\frac{\Delta m_{21}^2 L}{2 E}\right)-\cos 2 \theta_{23}^{} + 15 \right\} \;,  \\
   		Q_{\theta_{23}^{}}^{\nu_\mu^{}} &\simeq &  2 \cos 4 \theta_{23}^{} \sin ^2\left(\frac{\Delta m_{31}^2 L}{2 E}\right)+\cos \left(\frac{\Delta m_{31}^2 L}{E}\right)-8 \cos \left(\frac{\Delta m_{31}^2 L}{2 E}\right)+7 \nonumber \\
   		&& + 8 s^2_{13} \sin^2\left(\frac{\Delta m_{31}^2 L}{4 E}\right) \left[ \cos 2 \theta_{23}^{} - 2 \cos 4 \theta_{23}^{} \cos^2\left(\frac{\Delta m_{31}^2 L}{4 E}\right)+\cos \left(\frac{\Delta m_{31}^2 L}{2 E}\right)-2\right] \nonumber \\
   		&& + 2 c^2_{12} \frac{ \Delta m_{21}^2 L  }{E} \left[\sin^2 2 \theta_{23}^{} \sin \left(\frac{\Delta m_{31}^2 L}{E}\right)-2 \sin \left(\frac{\Delta m_{31}^2 L}{2 E}\right)\right] \;.
   	\end{eqnarray}
	To obtain the analytical results above, we have performed series expansions of the exact formulas in terms of $\sin^2\theta_{13}^{}$ and $\Delta m_{21}^2/\Delta m_{31}^2$. The corresponding results are plotted in the second and third rows of Fig.~\ref{fig:hnumu}, in which both the exact numerical results and the approximate analytical results are given. One can notice that the values of the QFI on neutrino mass-squared differences increase with $L/E$, while those on mixing angles and the {\it CP}-violating phase oscillate with $L/E$. Note that the range of $L/E$ in the plot of $Q_{\theta_{12}^{}}^{\nu_\mu^{}}$ has been extended to $40\,000~{\rm km}/{\rm GeV}$ to manifest its oscillation behavior. Since the next-generation long-baseline accelerator neutrino oscillation experiments are dedicated to measuring $\delta_{\rm CP}^{}$, we also indicate the corresponding values of $Q_{\delta_{\rm CP}^{}}^{\nu_\mu^{}}$ for DUNE (blue triangle), T2HK (blue inverted triangle) and ESSnuSB (blue square).
	
	\begin{figure}[t]
		\centering
		\includegraphics[scale=0.28]{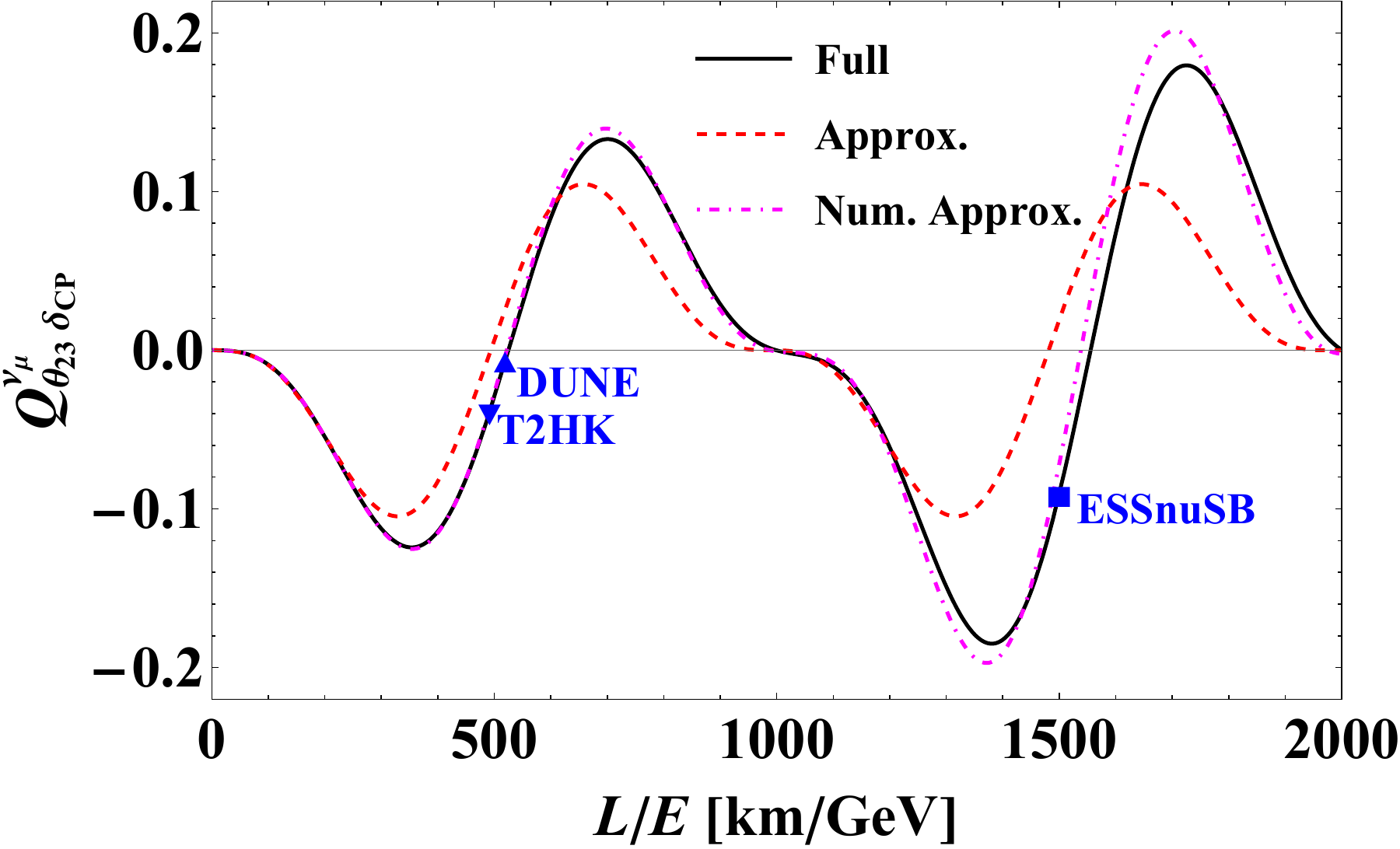}
		\includegraphics[scale=0.28]{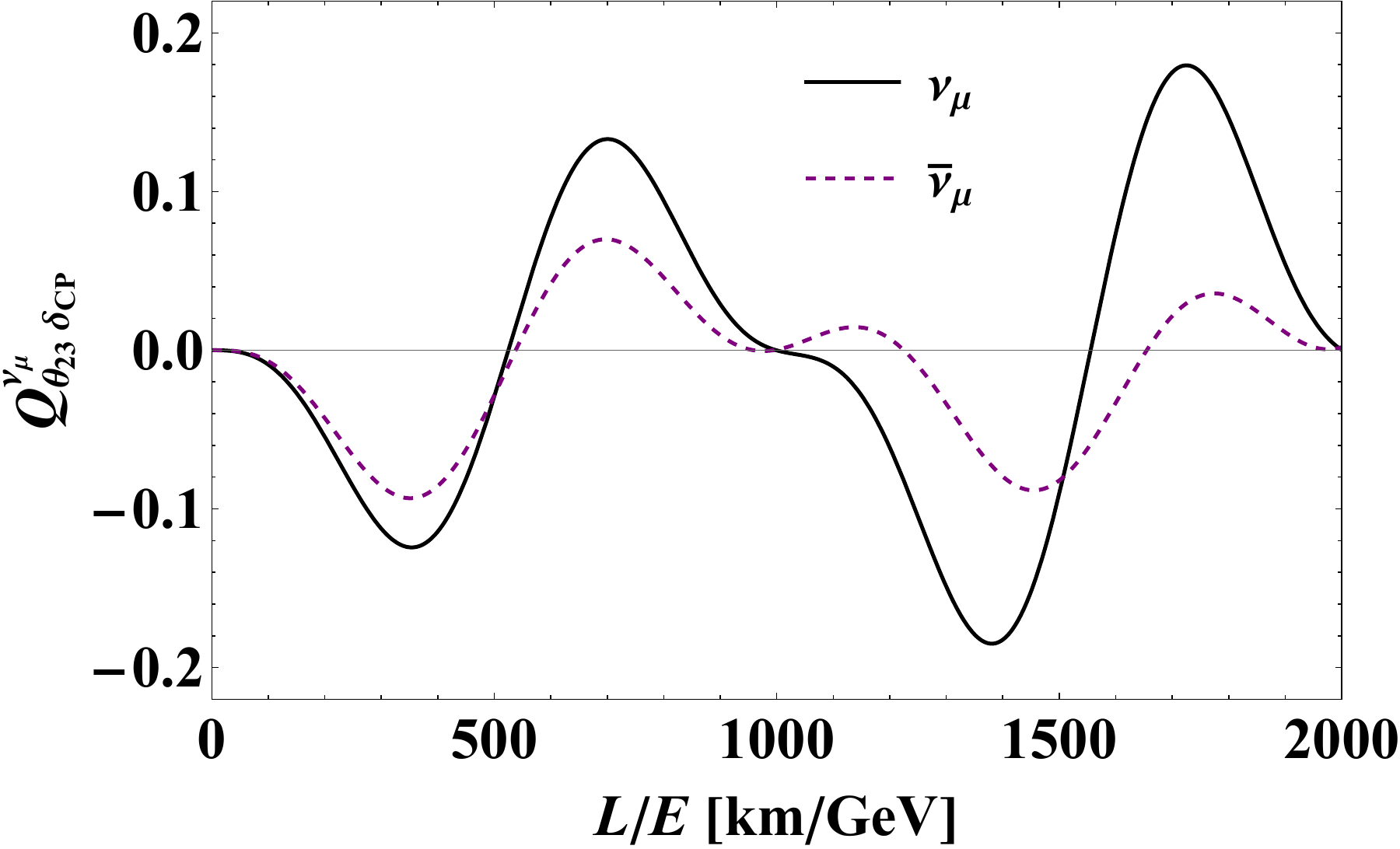}
		\vspace{-0.1cm}
		\caption{The off-diagonal elements $Q_{\theta_{23}^{} \delta_{\rm CP}^{}}^{\nu_\mu^{}}$ of the QFIM. The full numerical result is plotted as the black solid curve, while the red dashed curve shows the approximation given by Eq.~(\ref{eq:off_approx}). The magenta dot-dashed curve represents the numerical results with series expansion up to second order in $\sin^2\theta_{13}^{}$ and $\Delta m_{21}^2/\Delta m_{31}^2$. 
		The right panel shows the corresponding values of $Q_{\theta_{23}^{} \delta_{\rm CP}^{}}^{}$ for muon neutrinos and antineutrinos in the black solid and purple dashed curves, respectively.}
		\label{fig:htheta23deltanumu}
	\end{figure}
	
	The off-diagonal element $Q_{\theta_{23}^{} \delta_{\rm CP}^{}}^{\nu_\mu^{}}$ is plotted in Fig.~\ref{fig:htheta23deltanumu}, where we also show the approximate result at the leading order of $\sin^2\theta_{13}^{}$ and $\Delta m_{21}^2 / \Delta m_{31}^2$, i.e., 
   	\begin{eqnarray}
   		\label{eq:off_approx}
   		Q_{\theta_{23}^{} \delta_{\rm CP}^{}}^{\nu_\mu^{}} \simeq -64 s^2_{13} c^4_{13} s^3_{23} c_{23}^{} \sin ^3\left(\frac{\Delta m_{31}^2 L}{4 E}\right) \cos \left(\frac{\Delta m_{31}^2 L}{4 E}\right) \;.
   	\end{eqnarray}
	In the left panel of Fig.~\ref{fig:htheta23deltanumu}, apart from the full numerical result plotted as the black solid curve, we present the approximate result in Eq.~(\ref{eq:off_approx}) as the red dashed curve. In addition, the magenta dot-dashed curve represents the value of the off-diagonal QFI up to second order in both $\sin^2\theta_{13}^{}$ and $\Delta m_{21}^2/\Delta m_{31}^2$. Only at this second order can the analytical results reasonably reproduce the full numerical ones. However, the analytical formulas become very complicated and are hardly useful.   
	
	Finally, we emphasize that for antineutrinos, all results should be changed by replacing $\delta_{\rm CP}^{} \to -\delta_{\rm CP}^{}$. However, none of our approximate results involve $\delta_{\rm CP}^{}$ or $\sin\delta_{\rm CP}^{}$ at the given orders. Therefore, the differences in the QFI between neutrinos and antineutrinos appear as higher-order terms. 
	For example, in the right panel of Fig.~\ref{fig:htheta23deltanumu}, we plot the values of $Q_{\theta_{23}^{} \delta_{\rm CP}^{}}^{}$ for muon neutrinos (black solid curve) and muon antineutrinos (purple dashed curve). 
	Although no difference can be observed from the leading-order formula in Eq.~(\ref{eq:off_approx}), the numerical results indicate that their differences occur at higher order and can be very significant.

   	\subsection{QFIM and CR bounds}
   	
   	\label{subsec:QFIMandCR}
   	
	With all the analytical and numerical results obtained in previous subsections, we are ready to examine the precision of parameter measurements predicted by the QFIM. To be more explicit, we take the DUNE experiment as a concrete example and consider three specific cases for muon neutrinos: {\bf (i)} Only the diagonal elements of a QFIM; {\bf (ii)} A block-diagonal QFIM with a $4\times 4$ submatrix for three mixing angles and one {\it CP}-violating phase, and a $2\times 2$ submatrix for two independent neutrino mass-squared differences; {\bf (iii)} A full $6\times 6$ QFIM. 
	
	\subsubsection*{\underline{Case (i): A diagonal matrix}}
   	
    First, we ignore off-diagonal correlations and compute the CR bounds from the diagonal entries only.
    The diagonal QFIM for $\left\{\theta_{12}^{}, \theta_{13}^{}, \theta_{23}^{}, \delta_{\rm CP}^{}, \Delta m_{21}^2, \Delta m_{31}^2 \right\}$ at the DUNE experiment is
   	\begin{eqnarray}
   		Q = {\rm diag} 
   		\left\{ 0.0141, 7.52, 15.4, 0.156, 1.70\times 10^6~{\rm eV}^{-4}, 1.72\times 10^6~{\rm eV}^{-4} \right\} \;,
   	\end{eqnarray}
   	while the lower bound of the variance of any unbiased estimator is given by the CR inequality. For DUNE, we have the number of events $N={\cal O}(10^3)$, and the variances of six parameters are
   	\begin{eqnarray}
   		{\rm Var} \left[\theta_{12}^{}, \theta_{13}^{}, \theta_{23}^{}, \delta_{\rm CP}^{}, \Delta m_{21}^2, \Delta m_{31}^2 \right] & \gtrsim & {\rm diag} \left\{7.08 \times 10^{-2}, 1.33 \times 10^{-4}, 6.51 \times 10^{-5},  \right. \nonumber \\
   		&& \left. 6.40 \times 10^{-3},  5.88 \times 10^{-10}~{\rm eV}^4, 5.80 \times 10^{-10}~{\rm eV}^4 \right\} \;. \qquad
   	\end{eqnarray}
	In the most ideal case, the standard deviation of measuring three mixing angles and the {\it CP}-violating phase is no less than $\{15.2^\circ, 0.661^\circ, 0.462^\circ, 4.58^\circ\}$, while those on two mass-squared differences are $\{2.43\times 10^{-5}~{\rm eV}^2, 2.41\times 10^{-5}~{\rm eV}^2\}$.
	These numbers follow from the diagonal entries alone and completely neglect all correlations, so they should be regarded as single-parameter bounds rather than as simultaneously achievable uncertainties.

    To analyze the impact of the multiparameter estimation on the precision, we need to consider the complete $6\times 6$ QFIM. 
    The full QFIM can be defined as 
    \begin{eqnarray}
		Q = \begin{pmatrix}
			A & B \\ B^{\rm T} & C
		\end{pmatrix} \;,
	\end{eqnarray}
    where $A$ is the $4 \times 4$ QFIM for $\left\{ \theta_{12}^{},\theta_{13}^{},\theta_{23}^{},\delta_{\rm CP}^{} \right\}$ and $C$ is the $2 \times 2$ QFIM for $\{\Delta m_{21}^2, \Delta m_{31}^2\}$.
    The $4\times 2$ matrix $B$ describes the QFI between angles and mass-squared differences. 
    
    \subsubsection*{\underline{Case (ii): A block-diagonal matrix}}

	In this case, we simply focus on the information provided by $A$ and $C$.
    Numerically, we have
    \begin{eqnarray}
        \label{eq:A&C}
		A = 
		\begin{pmatrix} 
			0.0141 & 0.0785 & 0.0127 & 0.0340 \\
			0.0785 & 7.52 & -1.31 & 0.00629 \\
			0.0127 & -1.31 & 15.4 & -0.00612 \\
			0.0340 & 0.00629 & -0.00612 & 0.156
		\end{pmatrix} \;, \quad C=\begin{pmatrix}
		1.70  & -1.37  \\
		-1.37  & 1.72 
		\end{pmatrix} \times 10^6~{\rm eV}^{-4} \;.
	\end{eqnarray}
    To obtain the CR bounds on the corresponding parameters, one needs to calculate the inverse of those two matrices. 
    It is straightforward to verify that both $A$ and $C$ are invertible with
    \begin{eqnarray}
    \label{eq:A&C_inverse}
        A^{-1} = \begin{pmatrix}
			170 & -1.80 & -0.309 & -37.1 \\
			-1.80 & 0.154 & 0.0148 & 0.387 \\
			-0.309 & 0.0148 & 0.0666 & 0.0694 \\
			-37.1 & 0.387 & 0.0694 & 14.5 
		\end{pmatrix} \;, \quad C^{-1} = \begin{pmatrix}
			1.62 & 1.28 \\
			1.28 & 1.60 
		\end{pmatrix} \times 10^{-6}~{\rm eV}^4 \;. 
    \end{eqnarray}
    Then, we arrive at the covariance matrix from the CR inequality as
        \begin{eqnarray}
   		{\rm Cov}  \left[\theta_{12}^{},\theta_{13}^{},\theta_{23}^{},\delta_{\rm CP}^{}\right]  \gtrsim 10^{-3} A^{-1} \;, \quad {\rm Cov} \left[\Delta m_{21}^2, \Delta m_{31}^2\right] \gtrsim 10^{-3} C^{-1} \;.
   	\end{eqnarray}
   	The diagonal elements indicate that the lower bounds on the measurement uncertainties at the DUNE-like vacuum-benchmark configuration for the three mixing angles and the {\it CP}-violating phase could be no less than $\{23.6^\circ, 0.711^\circ, 0.468^\circ, 6.90^\circ \}$, while those for two neutrino mass-squared difference are $\left\{4.02\times 10^{-5}~{\rm eV}^2, 4.00\times 10^{-5}~{\rm eV}^2\right\}$. The variances of all six parameters are larger than those when taking into account only the diagonal elements. Moreover, the off-diagonal elements of the covariance matrices reflect the statistical correlation between any two relevant parameters, and their signs are physically meaningful. A positive (negative) off-diagonal element implies a positive (negative) correlation between those two parameters. In other words, an overestimation of one parameter tends to accompany an overestimation (underestimation) of the other.

	As pointed out in Sec.~\ref{sec:QFI}, the SLD--CR bound is attainable only when the SLD operators commute on average. We have evaluated the attainability condition in Eq.~(\ref{eq:commutator}) for all $15$ pairs of parameters at the DUNE-like configuration. Within the $4\times4$ block related to the matrix $A$ we obtain
	\begin{eqnarray}
		\mathrm{Im}\,Q^{\rm GT}_{ij} = \begin{pmatrix}
			0 & 0.045 & 0.085 & -0.003 \\
			-0.045 & 0 & -0.028 & -0.263 \\
			-0.085 & 0.028 & 0 & -0.047 \\
			0.003 & 0.263 & 0.047 & 0
		\end{pmatrix} \;, \qquad (i,j)=\left(\theta_{12}^{},\theta_{13}^{},\theta_{23}^{},\delta_{\rm CP}^{}\right) \;,
	\end{eqnarray}
	where $Q^{\rm GT}_{ij}\equiv \langle\partial_i^{}\psi|\partial_j^{}\psi\rangle - \langle\partial_i^{}\psi|\psi\rangle\langle\psi|\partial_j^{}\psi\rangle$. The imaginary parts are manifestly nonvanishing, in particular $\mathrm{Im}\,Q^{\rm GT}_{\theta_{23}^{},\delta_{\rm CP}^{}}=-0.047$, i.e., $\mathrm{Tr}[\rho\,[L_{\theta_{23}^{}},L_{\delta_{\rm CP}^{}}]]=-0.38\,\mathrm{i}\neq 0$. The optimal measurements for different mixing angles and $\delta_{\rm CP}^{}$ are therefore incompatible, and the bounds quoted in Cases~(i) and~(ii) are lower bounds that need not be attained simultaneously. By contrast, the two neutrino mass-squared differences are weakly compatible, with $\mathrm{Im}\,Q^{\rm GT}_{\Delta m_{21}^2,\Delta m_{31}^2}\approx 0$.

	For a genuinely multiparameter problem, the attainable asymptotic bound is more generally the Holevo CR bound~\cite{Ragy:2016ycy,Albarelli:2020pec}, which satisfies the hierarchy
	\begin{eqnarray}
		C_{\rm Holevo}^{} \geqslant C_{\rm SLD}^{} = \mathrm{Tr}\left[Q^{-1}\right] \;,
	\end{eqnarray}
	with equality if and only if $\mathrm{Im}\,Q^{\rm GT}_{ij}=0$ for all pairs. As a concrete illustration, for the two-parameter problem $\left(\theta_{23}^{},\delta_{\rm CP}^{}\right)$, the Holevo bound reads
	\begin{eqnarray}
		C_{\rm Holevo}^{} = \mathrm{Tr}\left[Q^{-1}\right] + \left\lVert Q^{-1}\, \mathfrak{U}\,Q^{-1}\right\rVert_1^{} \;,
	\end{eqnarray}
	with the mean Uhlmann curvature $\mathfrak{U}_{ij}^{}\equiv -({\rm i}/2)\,\mathrm{Tr}\left[\rho\,[L_i^{},L_j^{}]\right]$ and the trace norm $\lVert\cdot\rVert_1^{}$. Numerically, we find (in units of ${\rm rad}^{-2}$)
	\begin{eqnarray}
		C_{\rm SLD}^{}=6.46 \;, \qquad C_{\rm Holevo}^{}=6.62 \;, \qquad C_{\rm Holevo}^{}/C_{\rm SLD}^{}=1.024 \;,
	\end{eqnarray}
	so that the Holevo bound exceeds the SLD bound by about $2.4\%$, consistent with the small but nonvanishing incompatibility $\mathrm{Im}\,Q^{\rm GT}_{\theta_{23}^{},\delta_{\rm CP}^{}}=-0.047$.
    
    \subsubsection*{\underline{Case (iii): A full $6\times 6$ matrix}}

   	Finally, we consider the full $6\times 6$ QFIM for the parameter set $\left\{ \theta_{12}^{},\theta_{13}^{},\theta_{23}^{},\delta_{\rm CP}^{}, \Delta m_{21}^2, \Delta m_{31}^2 \right\}$.
    Using the numerically evaluated $4\times 2$ matrix $B$:
    \begin{eqnarray}
        \label{eq:B}
		B = \begin{pmatrix}
			-16.1 & 103 \\
			1390 & -3.49 \\
			253 & -31.5 \\
			227 & 9.89
		\end{pmatrix}~{\rm eV}^{-2} \;,
	\end{eqnarray}
    the full QFIM is given by
	\begin{eqnarray}
   		Q = 
   		\begin{pmatrix} 
   			14.1 & 78.5 & 12.7 & 34.0 & -1.61~\widetilde{\rm eV}^{-2} & 10.3~\widetilde{\rm eV}^{-2}
   			\\
   			78.5 & 7520 & -1310 & 6.29 & 139~\widetilde{\rm eV}^{-2} & -0.349~\widetilde{\rm eV}^{-2} \\
   			12.7 & -1310 & 15400 & -6.12 & 25.3~\widetilde{\rm eV}^{-2} & -3.15~\widetilde{\rm eV}^{-2} \\
   			34.0 & 6.29 & -6.12 & 156 & 22.7~\widetilde{\rm eV}^{-2} & 0.989~\widetilde{\rm eV}^{-2}
   			\\
   			-1.61~\widetilde{\rm eV}^{-2} & 139~\widetilde{\rm eV}^{-2} & 25.3~\widetilde{\rm eV}^{-2} & 22.7~\widetilde{\rm eV}^{-2} & 17.0~\widetilde{\rm eV}^{-4} &
   			-13.7~\widetilde{\rm eV}^{-4} \\
   			10.3~\widetilde{\rm eV}^{-2} & -0.349~\widetilde{\rm eV}^{-2} & -3.15~\widetilde{\rm eV}^{-2} & 0.989~\widetilde{\rm eV}^{-2} & -13.7~\widetilde{\rm eV}^{-4} &
   			17.2~\widetilde{\rm eV}^{-4}
   		\end{pmatrix} \times 10^{-3}  \;, \nonumber \\
   	\end{eqnarray}
	where we have defined the effective unit $\widetilde{\rm eV} \equiv 10^{-2}~{\rm eV}$ for convenience.
	In order to calculate the CR bound of the covariance matrix, we should first calculate the inverse matrix of $Q$.
    Since both $A$ and $C$ are invertible matrices, using the block-inversion formula, the inverse matrix $Q^{-1}$ reads
   	\begin{eqnarray}
   		\begin{pmatrix}
   			A & B \\ B^{\rm T} & C
   		\end{pmatrix}^{-1} &= &
   		\begin{pmatrix}
   			A^{-1} + A^{-1} B S_A^{-1} B^{\rm T} A^{-1} & \quad -A^{-1} B S_A^{-1} \\ -S_A^{-1} B^{\rm T} A^{-1} & \quad S_A^{-1}
   		\end{pmatrix}  \nonumber \\
   		&=& \begin{pmatrix}
   			S_C^{-1} & \quad -S_C^{-1} B C^{-1} \\
   			-C^{-1} B^{\rm T} S_C^{-1} & 
   			\quad C^{-1} + C^{-1} B^{\rm T} S_C^{-1} B C^{-1} \\ 
   		\end{pmatrix}  \;,
	\end{eqnarray}
	with $S_A^{} \equiv C - B^{\rm T} A^{-1} B$ and $S_C^{} \equiv A - B C^{-1} B^{\rm T}$.
	In particular, we have
	\begin{eqnarray}
		B^{\rm T} A^{-1} B = \begin{pmatrix}
		1.70  & -1.37  \\
		-1.37  & 1.72 
		\end{pmatrix} \times 10^6~{\rm eV}^{-4} \;, \quad S_A^{} \approx {\bf 0} \;,
	\end{eqnarray}
    and
    \begin{eqnarray}
        B C^{-1} B^{\rm T} &=& \begin{pmatrix} 
			0.0130 & 0.146 & 0.0222 & 0.0254 \\
			0.146 & 3.10 & 0.510 & 0.525 \\
			0.0222 & 0.510 & 0.0846 & 0.0863 \\
			0.0254 & 0.525 & 0.0863 & 0.0891
		\end{pmatrix} \;, \nonumber \\
        S_C^{} &=& \begin{pmatrix} 
		0.00112 & -0.0674 & -0.00948 & 0.00868 \\
		-0.0674 & 4.42 & -1.82 & -0.519 \\
		-0.00948 & -1.82 & 15.3 & -0.0924 \\
		0.00868 & -0.519 & -0.0924 & 0.0672
		\end{pmatrix} \;.
    \end{eqnarray}
    One should notice that all the matrix elements of $S_A^{}$ are rather small due to $C \simeq B^{\rm T} A^{-1} B$, and $S_C^{}$ is also a singular matrix with ${\rm det} \left( S_C^{} \right) \approx 0$.
    Consequently, we arrive at
    \begin{eqnarray}
		{\rm det} \begin{pmatrix}
			A & B \\ B^{\rm T} & C
		\end{pmatrix} = {\rm det} (A) {\rm det} \left( S_A^{} \right) = {\rm det} (C) {\rm det} \left( S_C^{} \right) \approx 0 \;,
	\end{eqnarray}
    i.e., $Q$ is singular with exactly two vanishing eigenvalues, as we now demonstrate. A normalized pure state of a three-flavor neutrino system can be written as the linear superposition of three neutrino mass eigenstates $|\psi\rangle = a^{}_1 |\nu^{}_1\rangle + a^{}_2 |\nu^{}_2\rangle + a^{}_3 |\nu^{}_3\rangle$, where $a^{}_i$ (for $i = 1, 2, 3$) are three complex numbers and thus in general contain six real parameters, but the normalization condition $\langle \psi|\psi\rangle = 1$ imposes one constraint $|a^{}_1|^2 + |a^{}_2|^2 + |a^{}_3|^2 = 1$ and the freedom of redefining the overall phase of $|\psi\rangle$ leads to the reduction of another one parameter. Therefore, for one normalized pure state, the rank of the QFIM is at most four. As a consequence, with six neutrino oscillation parameters, exactly two eigenvalues of $Q$ vanish identically, so that exactly two parameter combinations are unmeasurable from a single neutrino channel for any $L/E$.

    To make this explicit, we work in the dimensionless basis $\left\{\theta_{12}^{}, \theta_{13}^{}, \theta_{23}^{}, \delta_{\rm CP}^{}, \phi_{21}^{}, \phi_{31}^{}\right\}$ with $\phi_{ji}^{}\equiv \Delta m_{ji}^2 L/(2E)$, in which all QFIM entries are dimensionless and the eigenvalue problem is well posed. The complete eigenvalue spectrum reads
    \begin{eqnarray}
        \left\{\lambda_1^{},\dots,\lambda_6^{}\right\} = \left\{0,\ 0,\ 0.275,\ 1.69,\ 7.48,\ 15.6\right\} \;,
    \end{eqnarray}
    confirming that ${\rm rank}(Q)=4$. The two null eigenvectors (in the ordering $\{\theta_{12}^{},\theta_{13}^{},\theta_{23}^{},\delta_{\rm CP}^{},\phi_{21}^{},\phi_{31}^{}\}$), which span the unmeasurable subspace, are
    \begin{eqnarray}
        v_{\rm null}^{(1)} &\simeq& \left(0.059,\ 0.083,\ 0.014,\ 0.651,\ -0.585,\ -0.472\right) \;, \nonumber \\
        v_{\rm null}^{(2)} &\simeq& \left(-0.979,\ 0.007,\ 0.001,\ 0.180,\ 0.026,\ 0.096\right) \;.
    \end{eqnarray}
    The first is a genuine degeneracy relating $\delta_{\rm CP}^{}$ to the two mass-squared differences, while the second is dominated by $\theta_{12}^{}$. The explicit null parameter combination corresponding to the former reads
    \begin{eqnarray}
        v_{\rm null}^{} = 0.059\,\theta_{12}^{} + 0.083\,\theta_{13}^{} + 0.014\,\theta_{23}^{} + 0.651\,\delta_{\rm CP}^{} - 0.585\,\phi_{21}^{} - 0.472\,\phi_{31}^{} \;,
    \end{eqnarray}
    with $Q\,v_{\rm null}^{}\approx 0$. Since the two vanishing eigenvalues are degenerate, the null space is two-dimensional and the quoted vectors constitute one representative orthonormal basis thereof; the robust features are that one direction is dominated by $\theta_{12}^{}$, while the other is a $\delta_{\rm CP}^{}$--$\Delta m^2$ degeneracy. Owing to the exact vanishing eigenvalues, the condition number of $Q$ diverges [numerically $\kappa(Q)\sim 10^{17}$], and the six singular values in the dimensionless basis are $\{15.6,\ 7.48,\ 1.69,\ 0.275,\ \approx 0,\ \approx 0\}$. We have also verified numerically that ${\rm rank}(Q)=4$ for $L/E$ ranging from $50$ to $10^{4}~{\rm km}/{\rm GeV}$, so the singularity is a generic property of the single-channel (neutrino-only) estimation problem rather than an artifact of the DUNE value of $L/E$. Physically, the two flat directions mix $\theta_{12}^{}$ with $\delta_{\rm CP}^{}$ and the two mass-squared differences; their existence reflects both the weak imprint of the $\Delta m^2_{21}$-driven oscillations on a muon-neutrino beam (the diagonal entry $Q_{\theta_{12}^{}}^{\nu_\mu^{}}\approx 0.014$ is the smallest among the mixing angles) and a genuine degeneracy relating $\delta_{\rm CP}^{}$ to the mass-squared differences.

    The strict rank deficiency is resolved once neutrino and antineutrino data are combined. The antineutrino state is obtained by $\delta_{\rm CP}^{}\to -\delta_{\rm CP}^{}$, and the combined QFIM $Q_\nu^{}+Q_{\bar\nu}^{}$ (for equal statistics) has the eigenvalue spectrum
    \begin{eqnarray}
        \left\{3.0\times 10^{-5},\ 0.073,\ 0.64,\ 3.57,\ 14.6,\ 31.2\right\} \;,
    \end{eqnarray}
    so that all six eigenvalues are positive and $Q_\nu^{}+Q_{\bar\nu}^{}$ is full rank. The condition number drops to $\kappa\sim 10^{6}$, but the smallest eigenvalue remains small, indicating that $\theta_{12}^{}$ is still only weakly determined by muon (anti)neutrino data alone; external input, e.g., from reactor neutrino experiments, is still required to determine the full set of parameters precisely.
    
    \section{Summary}
	\label{sec:sum}

    In this work, we apply QET to investigate the ultimate precision limits on the measurement of neutrino oscillation parameters in a systematic way. Our investigation not only addresses some conceptual issues, but also provides quantitative results relevant for current and future neutrino oscillation experiments.
    We begin by clarifying a subtle point thus far overlooked in the literature, namely, the value of the QFI on a given parameter can depend on the physical basis in which the quantum state is expressed, provided that the unitary transformation connecting the two sets of bases itself depends on the parameter of interest. 
    Taking the two-flavor neutrino oscillations as an explicit example, we demonstrate that the QFI on the mixing angle $\theta$ differs significantly between the mass basis and the flavor basis, while the QFI on the mass-squared difference $\Delta m_{21}^2$ remains intact.  This serves as a reminder that the choice of physical bases in the calculation of the QFI must reflect the detection scheme of quantum states.

    Since neutrinos are produced and detected as flavor eigenstates, we have derived analytical expressions of the diagonal and off-diagonal elements of the QFIM for both electron and muon neutrinos in the case of three-flavor neutrino oscillations. The approximate results are also obtained via series expansions in terms of small parameters $\sin^2\theta_{13}^{}$ and $\Delta m_{21}^2/ \Delta m_{31}^2$, which are shown to agree well with full numerical calculations. For the electron antineutrino beam relevant to reactor neutrino experiments, the maximum value of QFI on $\theta_{13}^{}$ implies a CR bound on its variance of ${\cal O}(10^{-4})$, while that on $\delta_{\rm CP}^{}$ is about tens of times smaller, indicating that the {\it CP}-violating phase is intrinsically more difficult to measure using electron neutrinos alone.  We then compute the QFI for the muon neutrino beam. It has been found that the QFI on mass-squared differences grows monotonically with $L^2/E^2$, and those on mixing angles and $\delta_{\rm CP}^{}$ oscillate along with $L/E$.

    Meanwhile, we carry out a systematic treatment of multiparameter estimation in neutrino oscillations using the full QFIM. Specifically, we evaluate off-diagonal elements of a QFIM, such as $Q_{\theta_{23}^{} \delta_{\rm CP}^{}}^{\nu_\mu^{}}$, which reveal non-negligible correlations between parameters at the quantum level. Using DUNE as a concrete example, we compare the CR bounds obtained from the diagonal elements alone with those derived from the full matrix. Our results demonstrate that neglecting off-diagonal correlations leads to an underestimation of the lower bounds on measurement precision. For instance, the bound on $\theta_{12}^{}$ degrades from $15.2^\circ$ to $23.6^\circ$ when correlations are included, and that on $\delta_{\rm CP}^{}$ from $4.58^\circ$ to $6.90^\circ$. We analyze the attainability of the SLD bound through $\mathrm{Tr}[\rho\,[L_i^{},L_j^{}]]$, finding that the optimal measurements for different mixing angles and $\delta_{\rm CP}^{}$ are incompatible, and we discuss the Holevo bound, which exceeds the SLD bound by about $2.4\%$ for the $(\theta_{23}^{},\delta_{\rm CP}^{})$ pair. We further study the spectrum of the full $6\times6$ QFIM and show that its rank is exactly four, since the pure three-flavor neutrino state is restricted to the four-dimensional space; a single neutrino channel therefore leaves exactly two parameter combinations unconstrained. Combining neutrino and antineutrino data restores full rank, although $\theta_{12}^{}$ remains weakly determined. Throughout, our results are obtained for vacuum neutrino oscillations and are intended as state-level quantum-information benchmarks, rather than as realistic experimental sensitivity projections.

    Going beyond the results and discussions presented in this work, one could further incorporate matter effects and detection efficiencies into the QFI framework, and apply QET to study beyond-Standard-Model scenarios, such as neutrino nonstandard interactions. These extensions will not only elucidate the fundamental limits of precision measurements in neutrino physics, but also deepen our knowledge on relevant concepts in QET. We hope that our work serves as a useful reference for theoretical and experimental efforts in future applications of QET to neutrino oscillations and particle physics in general.

	\section*{Acknowledgments}
	
    The authors would like to thank Neetu R.S.~Chundawat for useful discussions. This work was supported in part by the National Natural Science Foundation of China under Grant No.~12475113 and No.~12535007, by the CAS Project for Young Scientists in Basic Research (YSBR-099), and by the Scientific and Technological Innovation Program of IHEP under Grant No.~E55457U2. 
	
		
	\bibliographystyle{utcaps_mod}
	\bibliography{ref_QFI}

@article{Frugiuele:2026yeq,
    author = "Frugiuele, Claudia and Genoni, Marco G. and Ignoti, Michela and Paris, Matteo G. A.",
    title = "{Quantum estimation theory limits in neutrino oscillation experiments}",
    eprint = "2602.16534",
    archivePrefix = "arXiv",
    primaryClass = "hep-ph",
    doi = "10.1007/JHEP08(2026)218",
    journal = "JHEP",
    volume = "08",
    pages = "218",
    year = "2026"
}

@article{Banerjee:2024lih,
    author = "Banerjee, Rajrupa and Panda, Papia and Mohanta, Rukmani and Patra, Sudhanwa",
    title = "{Analysis of neutrino oscillation parameters in the light on quantum entanglement}",
    eprint = "2410.05727",
    archivePrefix = "arXiv",
    primaryClass = "hep-ph",
    doi = "10.1016/j.nuclphysb.2025.117263",
    journal = "Nucl. Phys. B",
    volume = "1022",
    pages = "117263",
    year = "2026"
}

@article{ParticleDataGroup:2026aaa,
    author = "Takahashi, F. and others",
    collaboration = "Particle Data Group",
    title = "{Review of Particle Physics}",
    doi = "10.1142/S0217751X26300115",
    journal = "Int. J. Mod. Phys. A",
    volume = "41",
    pages = "2630011",
    year = "2026"
}

@article{Pontecorvo:1957cp,
	author = "Pontecorvo, B.",
	title = "{Mesonium and Antimesonium}",
	journal = "Sov. Phys. JETP",
	volume = "6",
	pages = "429--431",
	year = "1958"
}

@article{Maki:1962mu,
	author = "Maki, Ziro and Nakagawa, Masami and Sakata, Shoichi",
	title = "{Remarks on the unified model of elementary particles}",
	doi = "10.1143/PTP.28.870",
	journal = "Prog. Theor. Phys.",
	volume = "28",
	pages = "870--880",
	year = "1962"
}

@article{Liu:2016can,
    author = "Liu, Jing and Chen, Jie and Jing, Xiao-Xing and Wang, Xiaoguang",
    title = "{Quantum Fisher information and symmetric logarithmic derivative via anti-commutators}",
    eprint = "1501.04290",
    archivePrefix = "arXiv",
    primaryClass = "quant-ph",
    doi = "10.1088/1751-8113/49/27/275302",
    journal = "J. Phys. A",
    volume = "49",
    pages = "275302",
    year = "2016"
}

@article{Paris:2008zgg,
    author = "Paris, Matteo G. A.",
    title = "{Quantum Estimation for Quantum Technology}",
    eprint = "0804.2981",
    archivePrefix = "arXiv",
    primaryClass = "quant-ph",
    doi = "10.1142/s0219749909004839",
    journal = "Int. J. Quant. Inf.",
    volume = "07",
    pages = "125--137",
    year = "2009"
}

@article{Yadav:2026lsx,
    author = "Yadav, Bhavna and Subba, Amir and Shi, Yu",
    title = "{Quantum Fisher Information Revealing Parameter Sensitivity in Long-Baseline Neutrino Experiments}",
    eprint = "2602.05221",
    archivePrefix = "arXiv",
    primaryClass = "hep-ph",
    month = "2",
    year = "2026"
}

@article{Chundawat:2026lcm,
	author = "Chundawat, Neetu Raj Singh and Delgadillo, Luis A. and Li, Yu-Feng",
	title = "{Leptonic CP Phase Determination from Fisher Information in NO$\nu$A and T2K}",
	eprint = "2605.30404",
	archivePrefix = "arXiv",
	primaryClass = "hep-ph",
	month = "5",
	year = "2026"
}

@article{Chundawat:2026jjd,
    author = "Chundawat, Neetu Raj Singh and Li, Yu-Feng",
    title = "{Information-Theoretic Gaps in Solar and Reactor Neutrino Oscillation Measurements}",
    eprint = "2602.07991",
    archivePrefix = "arXiv",
    primaryClass = "hep-ph",
    month = "2",
    year = "2026"
}

@article{JUNO:2015zny,
	author = "An, Fengpeng and others",
	collaboration = "JUNO",
	title = "{Neutrino Physics with JUNO}",
	eprint = "1507.05613",
	archivePrefix = "arXiv",
	primaryClass = "physics.ins-det",
	doi = "10.1088/0954-3899/43/3/030401",
	journal = "J. Phys. G",
	volume = "43",
	pages = "030401",
	year = "2016"
}

@article{DUNE:2020ypp,
	author = "Abi, Babak and others",
	collaboration = "DUNE",
	title = "{Deep Underground Neutrino Experiment (DUNE), Far Detector Technical Design Report, Volume II: DUNE Physics}",
	eprint = "2002.03005",
	archivePrefix = "arXiv",
	primaryClass = "hep-ex",
	reportNumber = "FERMILAB-PUB-20-025-ND, FERMILAB-DESIGN-2020-02",
	doi = "10.2172/1599307",
	month = "2",
	year = "2020"
}

@article{Hyper-Kamiokande:2018ofw,
	author = "Abe, K. and others",
	collaboration = "Hyper-Kamiokande",
	title = "{Hyper-Kamiokande Design Report}",
	eprint = "1805.04163",
	archivePrefix = "arXiv",
	primaryClass = "physics.ins-det",
	month = "5",
	year = "2018"
}

@article{Alekou:2022emd,
	author = "Alekou, A. and others",
	title = "{The European Spallation Source neutrino super-beam conceptual design report}",
	eprint = "2206.01208",
	archivePrefix = "arXiv",
	primaryClass = "hep-ex",
	doi = "10.1140/epjs/s11734-022-00664-w",
	journal = "Eur. Phys. J. ST",
	volume = "231",
	pages = "3779--3955",
	year = "2022",
	note = "[Erratum: Eur. Phys. J. ST {\bf 232}, 15--16 (2023)]"
}

@article{JUNO:2022mxj,
	author = "Abusleme, Angel and others",
	collaboration = "JUNO",
	title = "{Sub-percent precision measurement of neutrino oscillation parameters with JUNO}",
	eprint = "2204.13249",
	archivePrefix = "arXiv",
	primaryClass = "hep-ex",
	doi = "10.1088/1674-1137/ac8bc9",
	journal = "Chin. Phys. C",
	volume = "46",
	pages = "123001",
	year = "2022"
}

@article{Capozzi:2025wyn,
	author = "Capozzi, Francesco and Giar{\`e}, William and Lisi, Eligio and Marrone, Antonio and Melchiorri, Alessandro and Palazzo, Antonio",
	title = "{Neutrino masses and mixing: Entering the era of subpercent precision}",
	eprint = "2503.07752",
	archivePrefix = "arXiv",
	primaryClass = "hep-ph",
	doi = "10.1103/PhysRevD.111.093006",
	journal = "Phys. Rev. D",
	volume = "111",
	pages = "093006",
	year = "2025"
}

@article{JUNO:2025gmd,
	author = "Abusleme, Angel and others",
	collaboration = "JUNO",
	title = "{Measurement of reactor neutrino oscillation with the first JUNO data}",
	eprint = "2511.14593",
	archivePrefix = "arXiv",
	primaryClass = "hep-ex",
	doi = "10.1038/s41586-026-10538-z",
	journal = "Nature",
	volume = "654",
	pages = "343--348",
	year = "2026"
}

@article{Helstrom:1967ldp,
    author = "Helstrom, C. W.",
    title = "{Minimum mean-squared error of estimates in quantum statistics}",
    doi = "10.1016/0375-9601(67)90366-0",
    journal = "Phys. Lett. A",
    volume = "25",
    pages = "101--102",
    year = "1967"
}

@book{Cramer1946,
    author = "Cram{\'{e}}r, Harald",
    title = "{Mathematical methods of statistics}",
    isbn = "0-691-08004-6",
    publisher = "Princeton University Press",
    year = "1946"
}

@book{Rao1945,
    author = "Rao, C. Radhakrishna",
	title = "{Information and the Accuracy Attainable in the Estimation of Statistical Parameters, Breakthroughs in Statistics: Foundations and Basic Theory}",
	doi = "10.1007/978-1-4612-0919-5_16",
	pages = "235--247",
	publisher = "Springer",
	year = "1992"
}

@article{Helstrom:1969fri,
	author = "Helstrom, Carl W.",
	title = "{Quantum detection and estimation theory}",
	doi = "10.1007/BF01007479",
	journal = "J. Stat. Phys.",
	volume = "1",
	pages = "231--252",
	year = "1969"
}

@article{Ignoti:2025rxr,
    author = "Ignoti, Michela and Frugiuele, Claudia and Paris, Matteo G. A. and Genoni, Marco G.",
    title = "{Is the large uncertainty of $\delta_{CP}$ fundamentally encoded in the neutrino quantum state?}",
    eprint = "2511.20148",
    archivePrefix = "arXiv",
    primaryClass = "hep-ph",
    month = "11",
    year = "2025"
}

@article{Liu:2019xfr,
	author = "Liu, Jing and Yuan, Haidong and Lu, Xiao-Ming and Wang, Xiaoguang",
	title = "{Quantum Fisher information matrix and multiparameter estimation}",
	eprint = "1907.08037",
	archivePrefix = "arXiv",
	primaryClass = "quant-ph",
	doi = "10.1088/1751-8121/ab5d4d",
	journal = "J. Phys. A",
	volume = "53",
	pages = "023001",
	year = "2020"
}

@article{Nogueira:2016qsk,
	author = "Nogueira, Edson C. and de Souza, Gustavo and Varizi, Adalberto D. and Sampaio, Marcos D.",
	title = "{Quantum estimation in neutrino oscillations}",
	eprint = "1610.05388",
	archivePrefix = "arXiv",
	primaryClass = "quant-ph",
	doi = "10.1142/s0219749917500459",
	journal = "Int. J. Quant. Inf.",
	volume = "15",
	pages = "1750045",
	year = "2017"
}

@article{Farooq:2026eap,
	author = "Farooq, Baktiar Wasir and Singh Koranga, Bipin and Chatterjee, Aritro",
	title = "{Quantum Fisher Information as a Probe of Sterile Neutrino New Physics:Geometric Advantage of KM3NeT over IceCube}",
	eprint = "2604.01256",
	archivePrefix = "arXiv",
	primaryClass = "hep-ph",
	month = "4",
	year = "2026"
}

@article{DayaBay:2025ngb,
	author = "An, F. P. and others",
	collaboration = "Daya Bay",
	title = "{Comprehensive Measurement of the Reactor Antineutrino Spectrum and Flux at Daya Bay}",
	eprint = "2501.00746",
	archivePrefix = "arXiv",
	primaryClass = "nucl-ex",
	doi = "10.1103/PhysRevLett.134.201802",
	journal = "Phys. Rev. Lett.",
	volume = "134",
	pages = "201802",
	year = "2025"
}

@article{Braunstein:1994zz,
	author = "Braunstein, Samuel L. and Caves, Carlton M.",
	title = "{Statistical distance and the geometry of quantum states}",
	doi = "10.1103/PhysRevLett.72.3439",
	journal = "Phys. Rev. Lett.",
	volume = "72",
	pages = "3439--3443",
	year = "1994"
}

@article{Yuen:1973mjw,
	author = "Yuen, H. and Lax, M.",
	title = "{Multiple-parameter quantum estimation and measurement of nonselfadjoint observables}",
	doi = "10.1109/TIT.1973.1055103",
	journal = "IEEE Trans. Info. Theor.",
	volume = "19",
	pages = "740--750",
	year = "1973"
}

@article{Albarelli:2020pec,
	author = "Albarelli, Francesco and Barbieri, Marco and Genoni, Marco G. and Gianani, Ilaria",
	title = "{A perspective on multiparameter quantum metrology: From theoretical tools to applications in quantum imaging}",
	eprint = "1911.12067",
	archivePrefix = "arXiv",
	primaryClass = "quant-ph",
	doi = "10.1016/j.physleta.2020.126311",
	journal = "Phys. Lett. A",
	volume = "384",
	pages = "126311",
	year = "2020"
}

@article{Liu:2015zxo,
	author = "Liu, Jing and Jing, Xiao-Xing and Wang, Xiaoguang",
	title = "{Quantum metrology with unitary parametrization processes}",
	eprint = "1409.4057",
	archivePrefix = "arXiv",
	primaryClass = "quant-ph",
	doi = "10.1038/srep08565",
	journal = "Sci. Rep.",
	volume = "5",
	pages = "8565",
	year = "2015"
}

@article{Akhmedov:2004ny,
	author = "Akhmedov, Evgeny K. and Johansson, Robert and Lindner, Manfred and Ohlsson, Tommy and Schwetz, Thomas",
	title = "{Series expansions for three flavor neutrino oscillation probabilities in matter}",
	eprint = "hep-ph/0402175",
	archivePrefix = "arXiv",
	reportNumber = "TUM-HEP-542-04",
	doi = "10.1088/1126-6708/2004/04/078",
	journal = "JHEP",
	volume = "04",
	pages = "078",
	year = "2004"
}

@article{Esteban:2024eli,
	author = "Esteban, Ivan and Gonzalez-Garcia, M. C. and Maltoni, Michele and Martinez-Soler, Ivan and Pinheiro, Jo{\~a}o Paulo and Schwetz, Thomas",
	title = "{NuFit-6.0: updated global analysis of three-flavor neutrino oscillations}",
	eprint = "2410.05380",
	archivePrefix = "arXiv",
	primaryClass = "hep-ph",
	reportNumber = "IFT-UAM/CSIC-24-140, YITP-SB-2024-24, IPPP/24/64, IPPP/24/64, IFT-UAM/CSIC-24-140, YITP-SB-2024-24",
	doi = "10.1007/JHEP12(2024)216",
	journal = "JHEP",
	volume = "12",
	pages = "216",
	year = "2024"
}

@misc{NuFIT:6-1,
    author = {Esteban, Ivan and others},
    collaboration = {NuFIT},
    title = {NuFIT 6.1},
    year = {2025},
    howpublished = "\url{www.nu-fit.org}"
}

@article{Esteban:2026phq,
	author = "Esteban, Ivan and Gonzalez-Garcia, M. C. and Maltoni, Michele and Martinez-Soler, Ivan and Pinheiro, Jo{\~a}o Paulo and Schwetz, Thomas",
	title = "{Lessons from the first JUNO results}",
	eprint = "2601.09791",
	archivePrefix = "arXiv",
	primaryClass = "hep-ph",
	reportNumber = "IFT-UAM/CSIC-26-3, IPPP/26/03, YITP-SB-2026-02",
	doi = "10.1007/JHEP04(2026)089",
	journal = "JHEP",
	volume = "04",
	pages = "089",
	year = "2026"
}

@article{Ragy:2016ycy,
    author = "Ragy, Sammy and Jarzyna, Marcin and Demkowicz-Dobrza\'{n}ski, Rafa{\l}",
    title = "{Compatibility in multiparameter quantum metrology}",
    eprint = "1608.02634",
    archivePrefix = "arXiv",
    primaryClass = "quant-ph",
    doi = "10.1103/PhysRevA.94.052108",
    journal = "Phys. Rev. A",
    volume = "94",
    pages = "052108",
    year = "2016"
}

@article{Dixit:2026qin,
    author = "Dixit, Khushboo and Kundu, Ritam and Panda, Papia and Razzaque, Soebur and Sarkar, Ramita",
    title = "{Quantum Information as a New Lens for Precision Neutrino Physics}",
    eprint = "2606.31996",
    archivePrefix = "arXiv",
    primaryClass = "hep-ph",
    reportNumber = "IP/BBSR/2026-09",
    doi = "10.1007/JHEP11(2026)016",
    journal = "JHEP",
    volume = "11",
    pages = "016",
    year = "2026"
}
	
\end{document}